\documentclass[
aps, 
pre, 
amsmath, 
article, 
superscriptaddress, 
reprint,
10pt,
longbibliography,
]{revtex4-2}
\UseRawInputEncoding
\usepackage[utf8]{inputenc}
\usepackage[english]{babel}
\usepackage{graphicx}
\usepackage{amsfonts}
\usepackage{bm}
\usepackage{soul}
\usepackage{color}
\usepackage{physics}
\usepackage{yhmath}
\usepackage{upgreek}

\newcommand\thalf{\ensuremath{{\textstyle\frac{1}{2}}}}

\newcommand\ie{i.e.\ }

\providecommand\bnabla{\bm{\nabla}}
\providecommand\bcdot{\bm{\cdot}}

\providecommand\bu{\bm{u}}
\providecommand\bd{\bm{d}}
\providecommand\boldf{\bm{f}}
\providecommand\bx{\bm{x}}

\providecommand\bxhat{\bm{\widehat{x}}}
\providecommand\byhat{\bm{\widehat{y}}}
\providecommand\bzhat{\bm{\widehat{z}}}
\providecommand\bk{\bm{k}}
\providecommand\bp{\bm{p}}
\providecommand\bq{\bm{q}}
\providecommand\bK{\bm{K}}

\providecommand\bV{\bm{V}}
\providecommand\bU{\bm{U}}

\providecommand\calA{\mathcal{A}}
\providecommand\calN{\mathcal{N}}
\providecommand\calD{\mathcal{D}}
\providecommand\calF{\mathcal{F}}
\newcommand\TFR{\mbox{TFR}}

\newcommand\Real{\mbox{Re}} 
\newcommand\Imag{\mbox{Im}} 

\providecommand\gh{\widehat{g}}

\providecommand\ghs{\widehat{g}^*}

\providecommand\tp{t^\prime}

\providecommand\defeq{:=} 

\newcommand{\wht}[1]{\widehat{\widetilde{#1}}}

\newcommand{\changes}[1]{}

\begin{document}


\title{Spatiotemporal spectral transfers in fluid dynamics} 


\author{Avik Mondal}
\email[]{abmondal@umich.edu}
\affiliation{Department of Physics, University of Michigan, Ann Arbor, MI 48109, USA}
\thanks{Co-first author}
\author{Andrew J. Morten}
\thanks{Co-first author}
\affiliation{Department of Physics, University of Michigan, Ann Arbor, MI 48109, USA}
\author{Brian K. Arbic}
\affiliation{Department of Earth and Environmental Sciences, University of Michigan, Ann Arbor, MI 48109, USA}
\affiliation{Research School of Earth Sciences, Australian National University, Canberra, ACT 2601, Australia}
\author{Glenn R. Flierl}
\affiliation{Department of Earth, Atmospheric, and Planetary Sciences, Massachusetts Institute of Technology, Cambridge, MA 02139, USA}
\author{Robert B. Scott}
\affiliation{D\'epartement de Physique et Laboratoire des Math\'emathiques Bretagne Atlantique, Universit\'e de Bretagne Occidental, CNRS, Brest 29238, France}
\author{Joseph Skitka}
\affiliation{Department of Earth and Environmental Sciences, University of Michigan, Ann Arbor, MI 48109, USA}


\date{\today}

\begin{abstract}
 Motivated by previous work on kinetic energy cascades in the ocean, atmosphere, plasmas, and other fluids, we develop a spatiotemporal spectral transfer tool that can be used to study scales of variability in generalized dynamical systems. In particular, we use generalized time-frequency methods from signal analysis to broaden the applicability of frequency transfers from theoretical to practical fluids applications such as the study of observational data or simulation output. We also show that triad interactions in wavenumber used to study kinetic energy and enstrophy cascades can be generalized to study triad interactions in frequency or wavenumber-frequency. We study the effects of sweeping on the locality of frequency transfers and frequency triad interactions to better understand the locality of spatiotemporal frequency transfers. As an illustrative example, we use the spatiotemporal spectral transfer tool to study the results of a simulation of two-dimensional homogeneous isotropic turbulence. This simulated fluid is forced at a well-defined wavenumber and frequency with dissipation occurring at both large and small scales, making this one of the first studies of ``modulated turbulence" in two dimensions. Our results show that the spatiotemporal transfers we develop in this paper are robust to potential practical problems such as low sampling rates or nonstationarity in time series of interest. We anticipate that this method will be a useful tool in studying scales of spatiotemporal variability in a wide range of fluids applications as higher resolution observations and simulations of fluids become more widely available.  

\end{abstract}


\maketitle 



%
%

%



\section{Introduction}

The cascade model plays a fundamental role in turbulence theory. According to the model, conserved quantities such as energy or enstrophy in two-dimensions are transferred locally between nearby spatial scales starting from some initial spatial scale where energy is injected and ending at some faraway dissipation scale.
The range of scales over which the cascade takes place is called the inertial range, within which forcing and dissipation have negligible direct effect.
While energy and enstrophy transfers are known to not be perfectly local \cite{kraichnan67,
brasseur94, zhou93, zhou96, frisch78}, the cascade framework  has proven useful for predictions of  wavenumber spectrum $\mathcal{E}(k)$ of energy in the inertial range.  Wavenumber spectra predictions were derived by ~\citet{Kolmogorov41c, Kolmogorov41a, Obukhov41a, Obukhov41b, Onsager45} for three-dimensional turbulence and by ~\citet{kraichnan67, Leith68, Batchelor69} for two-dimensional turbulence, and verified experimentally (less so for the two-dimensional case) and numerically many times over.

An important component of the cascade picture is the spectral transfer, which we  shall refer to as the \emph{spatial} spectral transfer to avoid ambiguity.  The spatial spectral transfer quantifies the time rate of change of some spatial spectral quantity, such as  the wavenumber spectra of kinetic energy $\mathcal{E}(k,t)$ or enstrophy $\mathcal{Z}(k,t)$, due to terms in the equation of motion.
For example, a kinetic energy spectral budget could take the form
\begin{equation}
\partial_t\mathcal{E}(k,t) = \sum_n \mathcal{A}_n(k,t),\label{eq:gen_spatial_budget}
\end{equation}
where $t$ is time and $k$ is isotropic wavenumber.
In this paper we will refer to each term $\mathcal{A}_n(k,t)$ in Eq.~\ref{eq:gen_spatial_budget} as a spatial spectral transfer arising from a different (the n-th)
term in the governing equations. 
Spatial spectral transfers refer to terms that contribute to the change in spatial spectral density of the quantity of interest at specific wavenumbers. In Eq.~\ref{eq:gen_spatial_budget}, $\mathcal{A}_n(k,t)$ contributes to the change in spatial spectral density in kinetic energy at wavenumber $k$ and time $t$. 

Each term $\mathcal{A}_n(k,t)$  can be thought of as transferring energy into or out of $\mathcal{E}(k,t)$ at wavenumber $k$. For example, an external forcing term might transfer energy into the system, and a dissipation term might transfer energy out of the system in the form of heat.
 In the cascade picture, the spatial spectral transfer due to the nonlinear advection term  plays a central role, as it is the only term that can move energy between scales.
 Indeed, sometimes the terminology ``spectral transfer'' is reserved exclusively for the nonlinear advection term.
 However, we find it useful to think of each term $\mathcal{A}_n(k,t)$ as a spectral transfer, whether the energy transfer is into the system, out of the system, or between spatial scales. Because we compute general results that apply to all $\mathcal{A}_n$'s, we refer to all of them as transfers in this paper.




In contrast with spatial spectral transfers, temporal and spatiotemporal spectral transfers, which would quantify the transfer of energy in frequency or wavenumber-frequency space respectively, have had little success in terms of theoretical predictions. 
For example, an analogous derivation of spectral slopes in frequency space would be problematic, because nonlinear transfers between time scales are expected to be less local than between spatial scales if small-scale structures are swept by large-scale structures \cite{Tennekes75}.
However, while temporal and spatiotemporal spectral transfers may have limited theoretical applications,
they can still be used as practical diagnostics. 
\citet{ScottArbic2007} used spatial spectral transfers to identify and examine an inverse cascade of kinetic energy present in a simple two-layer, quasigeostrophic model. \citet{Arbic12} diagnosed temporal spectral transfers in the ocean using 
altimetric measurements of sea surface height,
output from a realistic general ocean circulation model,
and the output from a simple two-layer quasi-geostrophic model. \citet{Arbic14} diagnosed spatiotemporal spectral transfers using similar data sets. 
  Both studies were motivated by the possibility of linking the well-known nonlinear inverse cascade of energy towards larger spatial scales with low-frequency variability in the ocean.

\citet{Mueller2015} quantified spatiotemporal spectral transfers due to internal gravity waves in two global ocean models at different resolutions. They found that the internal gravity wave frequency spectrum in the higher resolution simulation agreed better with spectra calculated from observations. 
\citet{Pan2020} and \citet{skitka23} used spatiotemporal spectral transfers to study internal gravity wave energy transfers in high-resolution regional ocean models.
\citet{Serazin2018} applied spatiotemporal spectral transfers to the long-duration output from a global ocean model, finding that advective transfers move energy from high-frequency frontal Rossby waves towards lower-frequency westward-propagating mesoscale eddies.
\citet{morten2017} used spatiotemporal spectral transfers to study kinetic energy dynamics in a simulation of single-layer, shallow-water beta-plane quasi-geostrophic turbulence. 
More recently, spatiotemporal and temporal transfers have been used to study interactions between the ocean and the atmosphere. \citet{ORourke2018} compared spatiotemporal transfers due to advection with transfers due to wind stress in the oceanic component of a coupled ocean-atmosphere model. \citet{Martin2020} employed temporal transfers of kinetic energy in the atmosphere and the ocean to understand the sources of variability at different timescales of each fluid in a quasi-geostrophic, fully coupled ocean-atmosphere simulation. \citet{Martin2021} extended this framework by computing a temperature variance budget to diagnose the sources of variability in an idealized ocean-atmosphere system as a function of frequency. Keating and Diamond use spectral transfers to study the effects of nonlinear wave-wave interactions on diffusivity \cite{Keating2007} and resistivity \cite{Keating2008} in turbulent plasma flows. 

The purpose of this paper is to provide a theoretical framework to aid in the interpretation of temporal and spatiotemporal spectral transfers calculated from theory, simulation, or data. As simulations and observations increase in resolution, we anticipate that temporal and spatiotemporal transfers will be useful for understanding the impact of small spatial scales on the overall dynamics of fluid flows. Small scale features, such as eddies and frontal features, can be resolved by new simulations and satellite measurements ~\cite{torres22, frengeretal}. The interactions of such high resolution features with other scales in fluid dynamics is a burgeoning area of study. Spectral transfers are a natural method of studying such interactions. Papers such as~\citet{Mueller2015, Serazin2018, Pan2020} and ~\citet{skitka23} demonstrate their usefulness in studying phenomena pertaining to wave turbulence. We anticipate that the thermodynamic effects of short timescale and small length scale features on oceanic and atmospheric dynamics can be probed by spectral transfers, as shown by ~\citet{ORourke2018} and ~\citet{Martin2021}.

Previous studies have derived spatiotemporal spectral transfers in theoretical or numerical contexts. In related work, \citet{Chiu1970} derived kinetic energy spectral equations in the frequency domain for large-scale atmospheric motions.  \citet{sheng1990a,sheng1990b} independently derived similar equations and applied them to global atmospheric simulations.   \citet{Elipot2009} also derived spectral equations to study the frequency components that dominate the wind energy input into the Ekman layer in the Southern Ocean.
All of these studies interpret the frequency spectra diagnostic as a balance among terms. For example, \citet{Chiu1970} notes that while the spectral budget in wavenumber space deals with the time rate of change of the spectrum,
 the analogous equation in frequency space contains no time rate of change and is therefore simply a balance among various co-spectra.
However, we show that a time rate of change can indeed be incorporated into the spectral budget in the frequency domain,
resulting in an improved interpretation. 
 
 To further aid in the interpretation of temporal and spatiotemporal spectral transfers, we investigate the effect of a mean flow.
 When there is a mean flow, one may invoke Taylor's hypothesis \citep{Taylor38,Lumley1965,Hill1996,Dosio2005,Bahraminasab2008,DelAlamo2009,Moin2009, Matthaeus2010, Servidio2011}, which states that small-scale structures with long lifetimes are advected without significant distortion by the large-scale, fast mean flow. In other words, at sufficiently small scales the turbulent structures appear to translate uniformly at the sufficiently high mean velocity $\bU$. 
 Under this assumption, there is a mapping between spatial and temporal Fourier modes given by $\omega = \bk\cdot\bU$, where $\omega$ is frequency. This assumption is used to study the behavior of a variety of turbulent fluids at different time scales; examples include ocean currents, convective currents in the atmosphere, and plasma flows from the Sun. In this paper, we calculate this mapping in the framework of spatiotemporal transfers by approximating a mean flow with a Galilean transformation of a fluid's velocity field and comparing this boosted velocity field to a reference velocity field. 

 Analysis of Taylor's hypothesis brings to light questions about the locality of spatiotemporal transfers as a result of a mean flow. To further study this locality, we introduce spatiotemporal triad interactions, which are a generalization of the wavenumber triad interactions introduced by \citet{kraichnan67}. We also discuss other uses and properties of these spatiotemporal triad interactions.






To illustrate the use of spatiotemporal spectral transfers, we apply the diagnostic to a simple fluid system:  a modified version of the incompressible two-dimensional Navier--Stokes equation~\citep[for reviews of two-dimensional turbulence see][]{ Tabeling02,Kellay02,Kraichnan80, Boffetta12}. 
Motivated by the utility of diagnosing  \emph{spatial} spectral transfers in systems where energy and enstrophy are injected within a narrow range of wavenumber ~\citep[recent examples include][]{Danilov2001,Chen03,Babiano2005,Chen06,Boffetta07,Xiao09,Dritschel2009}, we  choose to study the \emph{spatiotemporal} spectral transfers in a system where energy and enstrophy are injected within a narrow range of wavenumber \emph{and frequency}.  
As such, we apply a forcing that is approximately sinusoidal in space and time.
  
The effect of a sinusoidal modulated forcing has been investigated for three-dimensional turbulence ~\citep{Lohse00,Hooghoudt01,Heydt03,Cadot03,Kuczaj06,Kuczaj08}. In the three-dimensional case, resonant frequencies were predicted using the static structure functions of~\citet{Effinger87}.
To our knowledge no analogous result has been calculated for two-dimensional turbulence, which is complicated by the dual cascade of energy and enstrophy.  
%
The motivation for studying two-dimensional turbulence is to connect our results with prior studies of spatiotemporal spectral transfers calculated using oceanic data and output of realistic ocean models ~\citep{Arbic12,Arbic14}.
In those studies, for example, one region experienced an energy transfer to smaller time scales, an unexpected result that may be partly explained by the idealized numerical investigations used here.

A second motivation for the numerical investigation is to test the accuracy of the temporal spectral transfers as a diagnostic for datasets with limited temporal resolution or duration. Researchers studying large oceanic datasets often have to contend with both problems. For example, satellite altimeter measurements of sea surface height used to calculate spectral transfers suffer from both limited temporal resolution and duration, such that the data cannot resolve some relevant dynamical time scales. Because our numerical simulations produce data that resolve all dynamical time scales, we can comprehensively study the effects of limited record duration, limited temporal resolution, and temporal detrending in a way that is not necessarily possible with realistic data.  Our investigation shows that temporal spectral transfers can be relatively accurate over a range of frequencies under certain conditions even when the data set is severely limited.

In Sec.~\ref{sec:general}, we derive temporal and spatiotemporal spectral transfers for a general equation of motion. 
Because data sets are typically detrended in practical applications,
we also show how a detrending operation may be incorporated into the spectral transfers. If the correct detrending operation is applied, the spectral budget remains exact.

In Sec.~\ref{sec:fluids}, we assume that the equation of motion is two- or three-dimensional forced-dissipated Navier-Stokes, which has a nonlinear advection term. We derive temporal and spatiotemporal triad interactions and show how a mean flow 
 affects the spectral transfers and triad interactions. 

In Sec.~\ref{sec:numerics}, we calculate spatial, temporal, and spatiotemporal transfers using the output of a simulation of the incompressible two-dimensional Navier-Stokes equation. This simulation is a simplified version of the one used in \citet{morten2017}. We begin Sec.~\ref{sec:numerics} by introducing the details of this simplified simulation. We then diagnose the spatial, temporal, and spatiotemporal spectral transfers and fluxes for specific simulation outputs. We show how various limitations of the data or simulation output may affect the diagnosis of the transfers. Such limitations include poor temporal resolution, inadequate duration of the dataset, and the existence of a trend.  We show that for our two-dimensional simulations, the temporal transfers and fluxes are fairly robust when applied to data with the above limitations. We also show how the temporal spectral transfers change in time starting with a fluid at rest, diagnose the effects of detrending, and look for evidence of sweeping in the spatiotemporal spectral transfers. 

\subsection{Notation}

Throughout this paper, we employ a useful notational convention: a single function name may be used more than once to refer to several different functions,
with each unique function distinguished by the dimensions and units of the input parameters.
For example, $\calA_n(\bk,t)$, $\calA_n(k,t)$ and $\calA_n(k)$ are three different spatial spectral transfers. Similarly, we will later derive three different spatiotemporal and temporal spectral transfers:
$\calA_n(\bk, \omega, \tau)$, $\calA_n(k, \omega, \tau)$, and $\calA_n(\omega, \tau)$.
The distinction implied by the notation is that $\calA_n(\omega)$ is $\calA_n(\bk, \omega)$ with the $\bk$ dependence integrated out. The angular dependence is integrated out of $\calA_n(\bk, \omega, \tau)$ to get $\calA_n(k, \omega, \tau)$. Spectral energy densities $\mathcal{E}(\cdot)$ and spectral fluxes $\Pi_{\calA_n}^{>}(\cdot)$ will also follow the same notation.
This convention greatly reduces the number of symbols that the reader would otherwise need to remember. However, sometimes we want to avoid repeatedly writing ubiquitous parameters. In this case, we will introduce a function with parameters after a semicolon but then neglect to write these parameters in future uses of this function. This does not imply that the function's dependence on the neglected parameter is integrated out. We simply leave them out for terser notation.  For example, $\mathcal{E}(\bk,\omega, \tau; T)$  is the same function as $\mathcal{E}(\bk,\omega, \tau)$, but with the dependence on the parameter $T$  made explicit. Additionally, we will sometimes consider functions and their corresponding transforms (i.e various types of Fourier transforms or filtered functions). For such functions, we will use diacritics such as tildes, overbars, carons, and circumflexes depending on which transform we are applying to a function. We avoid such diacritics with spectral transfers $\mathcal{A}$, spectral fluxes $\Pi_{\calA_n}^{>}$, and spectral energy densities $\mathcal{E}$ because we only consider these in spectral space. We will denote multiple transformations to the same function with multiple diacritics.


\section{Spectral transfers for general system}\label{sec:general}

We derive temporal and spatiotemporal spectral transfers in a general way so that they may be used in a wide variety of fluid dynamical applications.
Accordingly, we begin with the most general equation of motion possible:
\begin{equation}
\partial_t g(\vb{x},t) = f(\vb{x},t).
\label{eq:general_eom}
\end{equation}
Above, $g$ is a general scalar field and $f$ gives the rate at which $g$ changes at any given point in space and time. $g$ can be used to refer to the components of vector fields. The function $f$ should be considered a finite sum of terms, such as forcing, dissipation, and nonlinear advection. It will be useful to write this explicitly as
\begin{equation}
f(\bx,t) = \sum\limits_{n} A_{n}(\bx,t),
\label{eq: forcings}
\end{equation}
where each $A_{n}$ contributes to the time derivative of $g$. 
We derive temporal and spatiotemporal spectral transfers with Eq.~\ref{eq:general_eom} as the starting point. Later on, we will specifically consider two- and three-dimensional fluid dynamical equations that contain a nonlinear advection term. 

It may help to motivate the use of Eq.~\ref{eq:general_eom} by considering some examples.
We could start with the incompressible Navier--Stokes equations with  general dissipation and forcing terms:
\begin{align}
\partial_t\bu(\bx,t)  = -(\bu \bcdot\bnabla)\bu -&\bnabla \Pi + \bd[\bu] + \boldf(\bx,t), \\ \bnabla\bcdot\bu&=0, \label{eq:advectioneq}
\end{align} 
where $\bu(\bx,t)$ is the Eulerian velocity field, $\Pi$ is pressure normalized by the density of water, $\bd[\cdot]$ is a fairly general linear dissipation operator
\begin{equation}
\bd[\bu] := -\sum_n \nu_n (-\nabla^2)^n\bu,
\end{equation}
with dissipation coefficients $\nu_n$,
and $\boldf(\bx,t)$ is an external force.
 We could start with the spatial Fourier transform of Eq.~\ref{eq:advectioneq}:
 \begin{align}
 \label{eq:generaleom3dk}
\partial_t\widetilde{\bu}(\bk,t) = -\widetilde{(\bu \bcdot\bnabla)\bu}  - &\widetilde{\bnabla \Pi} + \widetilde{\bd[\bu]} + \widetilde{\boldf}(\bk,t),\\  \bk\bcdot\widetilde{\bu}(\bk, t)&=0, 
\end{align} 
 where ``tilde" indicates a two- or three-dimensional spatial Fourier transform over a periodic domain. \citet{saltzman78} uses a version of Eq.~\ref{eq:generaleom3dk} to study large scale atmospheric turbulence.
 In this paper, Eq.~\ref{eq:generaleom3dk} will be the starting point for Sec.~\ref{sec:numerics}.
 If instead, we were interested in looking at individual terms in the multi-scale gradient expansion of ~\citet{Eyink06a, Eyink06b}, we could begin with the low-pass filtered equations \cite{Aluie_Eyink_09_partI, Aluie_Eyink_09_partII}:
 \begin{align}
\partial_t\overline{\bu}(\bx,t,\ell)  = -(\overline{\bu} \bcdot\bnabla)\overline{\bu} - \bnabla \overline{\Pi}& -\bnabla\bcdot\bm{\tau}+ \bd[\overline{\bu}] + \overline{\boldf}(\bx,t,\ell),\\\bnabla \cdot \overline{\bu} &= 0\label{eq:eyinkeq}
\end{align} 
where
\begin{equation}
\overline{A}(\bx,t,\ell):=\int d^{n}\bm{r}\; G_\ell(\bm{r})A(\bx+\bm{r}),
\end{equation}
$\bm{\tau}:= \overline{\bu\bu}-\overline{\bu}\;\overline{\bu}$, and $G_\ell$ is a low-pass filter kernel that keeps only length scales larger than $\ell$. 
We could also apply a spatial wavelet transform to Eq.~\ref{eq:advectioneq} or start with some other primitive fluid equation. The above list of examples of primitive equations and spatial transforms is not exhaustive.

\subsection{Spatial spectral transfers}\label{subsec:spatialtheory}

Spatial spectral transfers are relatively simple to derive. Combining Eq.~\ref{eq:general_eom} and  Eq.~\ref{eq: forcings} and taking the spatial Fourier transform of the resulting equation, we can write
\begin{equation}
\partial_t \widetilde{g}(\bk,t) = \sum_n \widetilde{A}_n(\bk,t).
\label{eq:general_eom_2}
\end{equation}
We can then multiply Eq.~\ref{eq:general_eom_2} by $\widetilde{g}^*(\bk,t)$ and take its real part to obtain the spectral budget,

\begin{equation}
  \partial_t\thalf|\widetilde{g}(\bk,t)|^2 = \sum_n \mathcal{A}_n(\bk,t),\label{eq:spatial_budget}
\end{equation}
where
\begin{equation}
  \mathcal{A}_n(\bk,t):=\Real\left[\widetilde{g}^*(\bk,t)\widetilde{A}_n(\bk,t)\right],
\end{equation}
and the superscript $*$ denotes a complex conjugate.
For each term $\widetilde{A}_n(\bk,t)$ in the equation of motion, $\mathcal{A}_n(\bk,t)$  is the corresponding spatial spectral transfer.
If $g$ were a vector rather than a scalar, then we would use a dot product with $\widetilde{g}^*(\bk,t)$. The imaginary part of the product of Eq.~\ref{eq:general_eom_2} and $\widetilde{g}^*{(\bk,t)}$ would tell us about the evolution of the phase of the $\widetilde{g}(\bk,t)$. We prove this in Appendix~\ref{app:phase} and present an example of how the imaginary part could be used.

As a concrete example, if the equation of motion were Eq.~\ref{eq:generaleom3dk} (as will be the case in Sec.~\ref{sec:fluids}), then we would have $\widetilde{g} = \widetilde{\bu}(\bk,t)$, and the spatial spectral budget would involve the time rate of change of
\begin{equation}
 \mathcal{E}(\bk,t) \defeq \thalf\left|\widetilde{\bu}(\bk,t)\right|^2,
\end{equation}
where $\mathcal{E}(\bk,t)$ is the energy in wavevector mode $\bk$ at time $t$. Motivated by this example,  we refer to 
$\thalf\left|\widetilde{g}(\bk,t)\right|^2$ as the ``energy.'' The spatial spectral budget Eq.~\ref{eq:spatial_budget}
tells us the time rate of change of energy in spatial mode $\bk$ due to each term in the equation of motion.
In other words, $\mathcal{A}_n(\bk,t)$ is the spatial spectral transfer corresponding to the term $\widetilde{A}_n(\bk,t)$ in the equation of motion.

\subsection{spatiotemporal spectral transfers}\label{subsec:spatialtemporaltheory}


One might expect to be able to derive temporal and spatiotemporal spectral transfers analogous to the derivation of spatial spectral transfers in Sec. ~\ref{subsec:spatialtheory}.
However, temporal spectral transfers are fundamentally different because of the 
temporal derivative in Eq.~\ref{eq:general_eom}.
 While the temporal derivative commutes with spatial detrending, spatial Fourier transforms, and other spatial transforms, the temporal derivative does not in general commute with temporal transforms. In this section, we discuss the difficulties with deriving spectral transfers in frequency space, many of which have been observed in previous works. Furthermore, we will demonstrate how to correctly calculate such spectral transfers.


Applying a temporal Fourier transform to Eq.~\ref{eq:general_eom}
results in the replacement of the time-derivative
by $i\omega$, giving
\begin{equation}
i\omega\check{\widetilde{g}}(\bk,\omega) = \check{\widetilde{f}}(\bk,\omega) = \sum_n \check{\widetilde{A}}_n(\bk,\omega).\label{eq:temporal_naive}
\end{equation}
In this paper, the ``caron'' operator, $\check{\hspace{0.1in}}$, denotes a temporal Fourier transform. 

Multiplying by the complex conjugate $\check{\widetilde{g}}^*(\bk,\omega)$ and taking the real part gives
\begin{equation}
0  = \sum_n \Real\left[ \check{\widetilde{g}}^*(\bk,\omega)\check{\widetilde{A_n}}(\bk,\omega) \right]\label{eq:temporal_balance}
\end{equation}
because multiplying any complex function by its complex conjugate and $i\omega$ results in an entirely imaginary function. Eq.~\ref{eq:temporal_balance} looks similar to the spatial spectral budget given by Eq.~\ref{eq:spatial_budget}, except that there is no longer a temporal derivative that leads to the interpretation of each term as a spectral transfer (\ie as a rate of change).  All that remains is a balance of terms. This is what \citet{sheng1990a} found when calculating atmospheric kinetic energy budgets. 

If we instead take the imaginary part we obtain
\begin{equation}
  \omega\left| \check{\widetilde{g}}(\bk,\omega)\right|^2 = \sum_n \Imag\left[ \check{\widetilde{g}}^*(\bk,\omega)\check{\widetilde{A_n}}(\bk,\omega) \right].\label{eq:imag_balance}
\end{equation}

The resulting Eq.~\ref{eq:imag_balance} resembles an energy budget and has been used in several other papers \cite{sheng1980, Elipot2009}.
However, the right-hand-side of Eq.~\ref{eq:imag_balance} does not say anything about the transfer of energy in spectral space. We will show in Appendix~\ref{app:phase} that it instead tells us about the phase dynamics of the time series of $g$. Our above usage of the Fourier transform, which resembles the derivation of a spatial-spectral transfers, cannot in general be used for practical applications for several reasons.

One such reason is that the Fourier transform is only defined for functions that decay to zero sufficiently rapidly as $t\rightarrow\pm\infty$. Time series from data or simulation do not necessarily explicitly decay to zero during the timescales of observation. While periodic boundary conditions can be used to study spatial scales of turbulent dynamics, no such approximation exists in time. This issue also applies to calculating the Fourier transforms of the autocorrelation function of $g(\bx,t)$ to exploit the Wiener-Khinchin theorem using the method of \citet{Chiu1970}. A Fourier transform of data that is not set to zero at the beginning and the end of the time window of observation will have ringing that will obscure the true spectral behavior of a function.

Furthermore, we are often interested in time series that are not statistically stationary in time. This usually manifests as a time signal having a time average over a finite window that changes significantly in time. Such time signals can be decomposed into a periodic component and a time-varying trend. The Fourier transform of a time series that includes a trend can be corrupted due to potential aperiodicities.

The above considerations motivate the need for a more careful derivation of spatiotemporal transfers. We turn to generalizations of the Fourier transform that work for statistically non-stationary, aperiodic time series, the so-called time-frequency representations ~\cite{hlawatsch08}. Time-frequency representations $\TFR_f(\omega,\tau)$ are functions derived from some time signal $f(t)$ of two variables: a frequency or scale $\omega$; and a ``central time'' $\tau$, which is typically the center of some time window of interest. Such time windows are usually defined by the range over which data of interest is available. At a fixed $\tau$, a time-frequency representation gives a transformed spectrum of a time series as a function of frequency or scale with respect to a chosen set of orthonormal basis functions. These transformations can solve the problems of aperiodicity and nonstationarity mentioned previously in this section. See~\citet{hlawatsch08} for details on general bilinear time-frequency representations. In this paper, we primarily focus on the spectrogram, which shows the frequency distribution of a signal as a function of central time and duration. Spectrograms are limited by the Heisenberg uncertainty principle, which prevents the precise colocalization of a function's frequency at a specific time \cite{hlawatsch08}. However, spectrograms allow for the most straightforward physical interpretation of the behavior of a function at different time and spatial scales. Furthermore, they are the most commonly used time-frequency representation in the previous fluids literature.  We briefly discuss the scalogram in Appendix~\ref{app:wavelet}. Scalograms characterize the distribution of scales in a signal. Rather than using Fourier transforms to calculate the signal's power at a certain time or length scale, they use a continuous wavelet transform. This transform replaces the complex exponential basis of Fourier transforms with a basis of wavelets, continuous functions with a characteristic scale and a customizable shape \cite{hlawatsch08}. They too are limited by the Heisenberg uncertainty principle and are straightforwardly physically interpretable. 

The spectrogram is the modulus squared of the short-time Fourier transform (STFT).  
For any function $f(t)$, the STFT is
%
\begin{align}
\widehat{f}(\omega;\tau, T) &:= \int_{\tau-T/2}^{\tau+T/2}\sigma(t-\tau;T)f(t) {e^{-i\omega (t-\alpha\tau)} dt},\label{eq:STFT}
\end{align}
where $\sigma(t-\tau;T)$ is a taper function of width $T$ centered at $t=\tau$. We use the circumflex to denote an STFT.
We assume $\sigma(t-\tau;T)$ to be identically zero outside of the range $\tau-T/2<t<\tau+T/2$, so one can also define the integration to be over the range $(-\infty,\infty)$ without changing the result of the integral.

\begin{figure}
\includegraphics[width=1\linewidth]{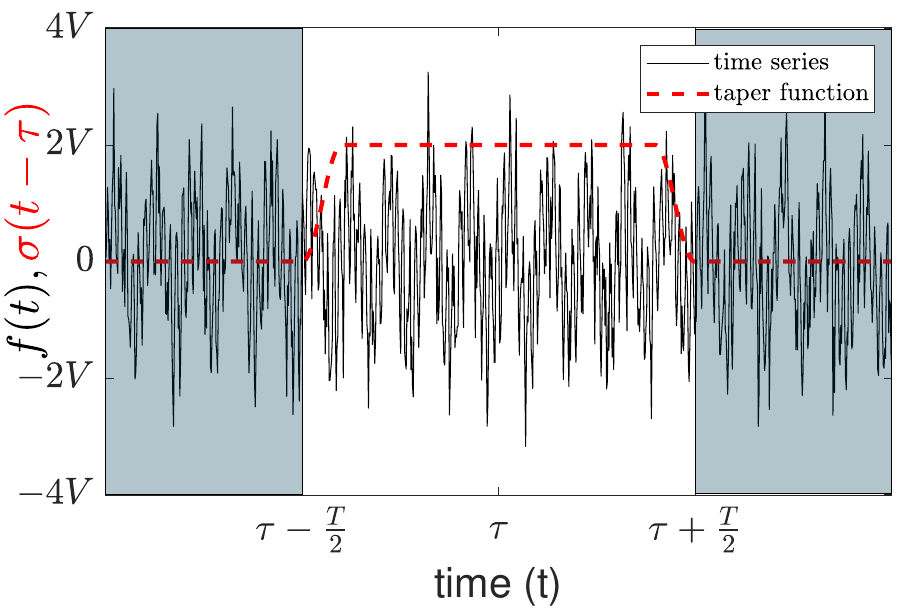}
\caption{Candidate time series $f(t)$ for STFT with taper function $\sigma (t)$ within a specified time window, denoted here in white. The grey-shaded region is not considered in an STFT calculation of $f(t)$ with chosen central time $\tau$ and window width $T$. The black solid curve corresponds to an artificial time series consisting of white noise with standard deviation $\frac{V}{2}$ and periodic functions of magnitude $V$. The red dotted curve corresponds to the taper function $\sigma (t)$  that guarantees that the time series decays to $0$ at the boundaries of the specified time window. We emphasize that while $T$ and $\tau$ are both time variables, $T$ sets a timescale while $\tau$ sets the central time location of the specified window.}
\label{fig: stftplot}
\end{figure}

The STFT is effectively a Fourier transform applied to a tapered signal with the central position $\tau$ of the taper function made explicit. Whenever the choice of taper function $\sigma$ must be made explicit, we use the notation $\widehat{f}[\sigma](\omega, \tau; T)$. We show an example of a potential time series with a Tukey (20\% taper) function as the taper function in Fig.~\ref{fig: stftplot}. While many candidate functions for $\sigma(t)$ are referred to as windows (Gaussian window, Tukey window, etc.), we refer to them as taper functions in this paper to avoid confusion with the time windows centered at $\tau$. Note that because we apply STFT's with a finite $T$, the case of ``no taper" is effectively the same as setting $\sigma(t)$ as a rectangular function:
\begin{equation}
\label{eqn:recttaper}
\sigma_R(t - \tau ; T) = \Theta\Big(t - \tau + \frac{T}{2}\Big) - \Theta\Big(t - \tau - \frac{T}{2}\Big).
\end{equation}
In the above equation, $\Theta$ is the Heaviside step function. In this paper, we do not analyze  the advantages of different tapers. In our numerical section, Sec. \ref{sec:numerics}, we apply the Tukey function shown in Fig.~\ref{fig: stftplot} to the signals. This taper reduces ringing without dramatically changing the properties of the signal of interest. Additionally, this is the taper function used quite frequently, particularly by our research group \cite{Arbic13, Arbic14, Martin2020, Martin2021, Hochet2020}. We therefore want to understand the effect of using this taper in a simplified numerical system. For other tapering functions and a detailed consideration of their properties, we point readers to the analysis in Chapter 13.4   of \citet{numericalrecipes}.

The parameter $\alpha$ in Eq.~\ref{eq:STFT} takes one of two values: $\alpha=0$ or $\alpha=1$. The standard definition of STFT has $\alpha=0$ with basis functions $e^{-i\omega t}$ centered at $t=0$, which corresponds to complex exponential functions that do not move with the time window. It will be useful to consider the $\alpha=1$ case, which has Fourier basis functions $e^{-i\omega(t-\tau)}$ centered at $t=\tau$, which corresponds to complex exponential functions that \textit{do} move with the time window.
The two cases differ only by a multiplicative factor of $e^{i\omega\tau}$ that commutes with the integral in Eq.~\ref{eq:STFT}. 


Applying the STFT to Eq.~\ref{eq:general_eom_2} gives
\begin{equation}
  \int_{-\infty}^{\infty}\sigma(t-\tau;T)\left(\partial_t \widetilde{g}(\bk,t)\right) e^{-i\omega (t-\alpha\tau)} dt 
  =  \sum_n \widehat{\widetilde{A_n}}(\bk,\omega,\tau).
\end{equation}
Integrating by parts and
transforming $t$-derivatives into $\tau$-derivatives gives
\begin{equation}
\partial_\tau \widehat{\widetilde{g}}(\bk,\omega,\tau) +(1-\alpha)i\omega\widehat{\widetilde{g}} = \widehat{\widetilde{f}}(\bk,\omega,\tau). \label{eq:tau_derivative_g}
\end{equation}
Multiplying Eq.~\ref{eq:tau_derivative_g} by $\frac{1}{2}\widehat{\widetilde{g}}^*$ and then adding the product of $\frac{1}{2} \widehat{\widetilde{g}}$ and the complex conjugate of Eq.~\ref{eq:tau_derivative_g} gives the equation for the $\tau$-derivative of the spectrogram of $g$:
\begin{equation}
  \thalf\partial_\tau|\widehat{\widetilde{g}}(\bk,\omega, \tau)|^2 = \Real[\widehat{\widetilde{g}}^*\widehat{\widetilde{f}}]. 
  \label{eqn:genergy1234}
\end{equation}
We also refer to the spectrogram of $g$ as an ``energy," since many readers may be most accustomed to the case where $g=\vb{u}$, in which case the spectrogram of $g$ is the kinetic energy. Expanding $f$ as a sum of terms gives the spatiotemporal spectral budget:
\begin{equation}
  \thalf\partial_\tau|\widehat{\widetilde{g}}(\bk,\omega, \tau)|^2 = \sum_n \mathcal{A}_n(\bk,\omega,\tau).\label{eq:temporal_budget_general}
\end{equation}
where
\begin{equation}
\label{eq:spatiotemporalspectraltransfers}
\mathcal{A}_n(\bk,\omega, \tau) := \Real[\widehat{\widetilde{g}}^*(\bk,\omega, \tau)\widehat{A_n}(\bk,\omega,\tau)].
\end{equation}
Unlike the balance obtained in Eq.~\ref{eq:temporal_balance}, the spectral budget Eq.~\ref{eq:temporal_budget_general} contains a $\tau$-derivative of the energy. This derivative is a natural consequence of calculating the time-frequency representation of a time derivative. In the above equations, we see the result of applying an STFT to a time derivative. An analogous $\tau$-derivative manifests if one instead decides to calculate the wavelet transform, as we demonstrate in Appendix~\ref{app:wavelet}. 

The $\tau$-derivative of the energy $\thalf|\widehat{\widetilde{g}}(\bk,\omega,\tau)|^2$ tells us how the energy in mode $(\bk,\omega)$ changes as the central time changes (\ie as the window of the STFT advances). The interpretation of the temporal spectral budget Eq.~\ref{eq:temporal_budget_general} is therefore analogous to that of the spatial-spectral budget Eq.~\ref{eq:spatial_budget} with the $t$-derivative replaced by a $\tau$-derivative. Eq.~\ref{eq:temporal_budget_general} can instead be derived using the autocorrelation function of a time series in the manner of \citet{Chiu1970} instead of the above method; we show this derivation in detail in Appendix~\ref{app:wkth}.For a statistically homogeneous time series with a large enough time window, the $\tau$-derivative will act as a residual. If all the appropriate terms are included in Eq.~\ref{eq:general_eom_2} and a large enough range of data is considered, the $\tau$-derivative should equate to zero. If there are stochastic terms in the equation of motion, then one would expect the average of the $\tau$-derivative over a sufficiently large ensemble would be zero. One may find that the $\tau$-derivative is non-zero if the chosen time window is too small to capture the long period contributions to the input signal; we will show examples of this in Figs.~\ref{fig:result_windowingsampling} and ~\ref{fig:result_detrending}. 

Eq.~\ref{eq:temporal_budget_general} allows one to define spatiotemporal spectral transfers, just as Eq.~\ref{eq:gen_spatial_budget} allows one to define spatial-spectral budgets. Previous research studied all the terms we define as spatiotemporal spectral transfers, not just nonlinear advection \cite{Hayashi1982, Elipot2009,sheng1990a, Martin2020, Martin2021, ORourke2018, Arbic14}. In contrast to previously defined spatiotemporal spectral transfers, our definition includes a dependence on time variables \cite{Hayashi1982, Elipot2009,sheng1990a}. This is necessary when considering spatiotemporal spectral transfers of time series that could be aperiodic.  The central time $\tau$ considered with a specific window width $T$ determines what subset of a time series is analyzed. Different values of $\tau$ can lead to different amplitudes at each frequency mode due to the presence of noise or nonstationarity in a time series. 

Window width $T$ affects spatiotemporal spectral transfer as well because of the Heisenberg uncertainty principle. Specifically, $T$ is inversely proportional to the width of the frequency bins for the spatiotemporal spectral transfers. Changing $T$ can change the amplitude of transfers in specific bins. This same effect is visible in other forms of the discrete Fourier transform. We will for the most part leave $T$ out of the notation for the sake of terseness. The use of the STFT on data or simulation output does introduce a dependence on the sampling rate of the time signal of interest. The sampling rate is inversely proportional to the maximum calculable frequency of the spatiotemporal spectral transfer. We will show the effect of $T$ and sampling rate on temporal spectral transfers in Fig.~\ref{fig:result_windowingsampling}. 

The above calculation is compatible with Welch's method if applied carefully \cite{welch1967}. With Welch's method, a power spectrum or a cross-spectrum is estimated by partitioning a time series into equally sized smaller time series. The periodograms of these smaller time series are then calculated and averaged to provide an estimator for the true spectrum of the time series. This is a powerful method for estimating power spectra of noisy time series if one can assume that the smaller time series are statistically identical. In fact, it can be useful to use the $\tau$-derivative to determine if using Welch's method is appropriate to use on a time series. If the value of this derivative calculated for each smaller time series is non-negligible, it is possible that the length of the smaller time series is too small to capture low-frequency phenomenon extant in the system of interest (See Fig.~\ref{fig:result_detrending}).This would be an example of broken ergodicity, where the statistical properties of the time series are changing in time\cite{nonergodic, palmer1982}. If this is the case, the spectra of the smaller time series cannot be used to calculate an average. In this context, we recommend only analyzing time series that are long enough to have the smallest possible $\tau$-derivative. Minimizing the $\tau$-derivative could be a useful method for deriving a minimum $T$ for any specific analysis. Alternatively, large $\tau$-derivatives could also be a sign of an imbalanced spectral budget. If this is the case, Welch's method will not improve the accuracy of estimated spectra. 

Importantly, Welch's method cannot replace tapering in the STFT. Tapering systematically ameliorates spurious spectral signals caused by aperiodicities. Welch's method will reduce the effects of noise in un-tapered periodograms, but cannot dependably fix inaccuracies caused by aperiodicities.

In the STFT defined in Eq.~\ref{eq:STFT}, we do not account for nonstationary functions whose spatial average changes in time. We dedicate the next subsection to this topic. We do not address methods for nonstationary functions with evolutionary spectra. Such functions have power spectra that change in time and have added complexity outside this paper's scope. The methods discussed in the next section cannot completely account for evolutionary spectra.




\subsection{Effects of detrending}

In practice, the standard method of dealing with nonstationarity is detrending \cite{Priestleybook}. Detrending subtracts a chosen function from a time series, derived from a best-fit of a function with the time series. In previously published papers by some of the authors, detrending consists of computing the linear trend of the time series and subtracting it from the time series \cite{ Martin2021, Martin2020, morten2017, Arbic12, Arbic14}. Removing this best-fit line removes a square wave or sawtooth waveform from the time series. This removes ringing artifacts from the Fourier transform that corrupt the entire frequency range of a time series. We emphasize that simply subtracting a linear trend is distinct from applying any type of low or high-pass filters which can be used to separate time series into ``slow" and ``fast" parts. Separating time scales does not entirely remove ringing effects because ringing corrupts the spectra at all frequencies.




In this section we show how to derive an exact spectral budget, giving the $\tau$ rate of change of the energy in the detrended signal:
\begin{equation}
\frac{1}{2}\partial_\tau|\widehat{\widetilde{g_{detrend}}}(\bk,\omega; \tau)|^2 = \Real[\widehat{\widetilde{g_{detrend}}}^*\widehat{\widetilde{f_{detrend}}}].
\label{eq:det_budget}
\end{equation}
Here, the subscript ``detrend'' indicates the application of an operator that removes the trend from a function.
Eq.~\ref{eq:det_budget} is not obvious and in fact depends on the details of how the detrending is formulated and incorporated with the STFT. In this section, we allow for detrending operations beyond subtracting a linear trend. We derive the requirements necessary for a generalized detrending operation to give a spectral budget of the form Eq.~\ref{eq:det_budget}.  

Given a function $h(t)$, we calculate its trend over a finite window of width $T$ centered at time $\tau$. Thus, the trend will depend on the position of the window, so it will have the form $h_{trend}(t,\tau)$.  For example, a linear trend  takes the form
\begin{equation}
h_{trend}(t,\tau) = c_0(\tau) + c_1(\tau)(t-\tau),
\end{equation}
where $c_0(\tau)$ and $c_1(\tau)$ are the fit coefficients, which depend on the position of the window. For more general trends, because we only calculate spectrograms for finite time windows, we can define a trend function as:
\begin{equation}
\label{eq:lls}
h_{trend}(t,\tau) := \sum_n c_n(\tau)\varphi_n(t-\tau)
\end{equation}
where the $\varphi_n(t-\tau)$ refer to members of an orthogonal basis that move with the position of the window. We use angle brackets to denote the inner product for basis functions $\varphi$:

\begin{equation}
 \braket{\varphi_n(t^\prime)}{\varphi_m^*(t^\prime)} _{t^\prime} \defeq \int_{-T/2}^{T/2}dt^\prime \varphi_n(t^\prime) \varphi_m^*(t^\prime).
\end{equation}
For simplicity, the above calculation considers only trends constructed from bases with unweighted inner products; for example, for polynomial trends of order $b$ one can use the first $b$ Legendre polynomials. Any family of orthogonal functions whose inner product space corresponds with the time window can be used. 

We use Eq. \ref{eq:lls} to define the detrending operation: 
\begin{equation}
\label{eqn: detop}
h_{detrend} \defeq h(t) - h_{trend}(t,\tau).
\end{equation}

Because it is commonly used, we consider in detail the linear least squares method for detrending.
The coefficients $c_n$ are determined by minimizing the squared residual as shown in Appendix~\ref{app:detrend}. The resulting best-fit coefficients are
\begin{equation}
\label{eq:c_n}
c_n(\tau) = \sum_m (M^{-1})_{nm} \braket{h(t^\prime+\tau)} {\varphi_m^*(t^\prime)}_{t^\prime},
\end{equation}
In Eq.~\ref{eq:c_n}, $M$ is an invertible matrix independent of $t$ and $\tau$, defined in Eq.~\ref{eqn:Mdef} in Appendix \ref{app:detrend}. Plugging Eq. \ref{eq:c_n} into Eq. \ref{eq:lls} gives the trend $h_{trend}(t,\tau)$ in terms of $f(t)$:

\begin{equation}
h_{trend}(t,\tau) = \sum_n \varphi_n(t-\tau)\sum_m (M^{-1})_{nm}\braket{ h(t^\prime+\tau)}{\varphi_m^*(t^\prime)}_{t^\prime}.
\label{eq:trend_explicit}
\end{equation}

With Eq.~\ref{eq:trend_explicit}, we can construct trends of any form that can be derived with linear least squares and an orthogonal basis with an unweighted inner product. To derive a detrended spectral budget, we can combine Eq. ~\ref{eq:general_eom} and Eq.~\ref{eq:det_budget} to get:
\begin{equation}
(\partial_t g)_{detrend}(\bx,t) = f_{detrend} (\bx,t).
\label{eq: general_eom_det}
\end{equation}

We separately calculate $(\partial_t g)_{trend}$, $\partial_t g_{trend}$ and $\partial_\tau g_{trend}$ in Appendix~\ref{app:detrend} and combine the results to obtain 
\begin{equation}
(\partial_t g)_{trend} = (\partial_\tau + \partial_t)g_{trend}(t,\tau),\label{eq:trendofgdot}
\end{equation}
from which it easily follows that
\begin{equation}
(\partial_t g)_{detrend} = (\partial_\tau + \partial_t)g_{detrend}(t,\tau).\label{eq:detofgdot}
\end{equation}
In other words,
applying the time-derivative \emph{before} removing the trend gives the same result as applying the operator  $(\partial_\tau + \partial_t)$ \emph{after} removing the trend.


In addition to satisfying Eq.~\ref{eq:detofgdot}, linear least squares satisfies linearity,
\begin{equation}
(f_1+f_2)_{detrend} = (f_1)_{detrend} + (f_2)_{detrend}, \label{eq:detrend_linear}
\end{equation}
which is useful because the spectral budget can remain a sum of terms before and after a linear least squares detrending operation.

To continue the derivation of the spectral budget, we would like to apply the STFT after the detrending operation.
However, some care is required in applying the STFT after detrending.
Whereas the standard STFT takes as input a function dependent only on the time variable $t$, after detrending we actually have a function dependent on $t$ and window time position $\tau$.
It is natural to equate the window position $\tau$ used by the detrending operation with the window position $\tau$ used by the STFT, which is why we have already used $\tau$ in both situations.
Given a function of the form $f(t,\tau)$ as input, we therefore define the generalized STFT as
\begin{align}
\mbox{STFT}_{f}(\omega,\tau) := \int_{\tau-T/2}^{\tau+T/2}\sigma_T(t-\tau)f(t,\tau) {e^{-i\omega (t-\alpha\tau)} dt}.\label{eq:STFTdetrend}
\end{align}
The change is subtle, but it is important to understand that the input to the STFT depends on the central time when we incorporate detrending.
We use the same notation as for the standard STFT, since the only change is in the form of the input function.

Combining Eqs.~\ref{eq:STFT}, ~\ref{eq:tau_derivative_g}, and ~\ref{eq:detofgdot}, we find that the generalized STFT of the detrended derivative of a function $g$ is
\begin{align}
  \nonumber \widehat{\left(\partial_tg\right)_{detrend}}(\omega,\tau) 
  =\partial_\tau \widehat{g_{detrend}}&(\omega,\tau) \: + \\ & \: \: \: \:  (1-\alpha)i\omega \widehat{g_{detrend}}(\omega,\tau).\label{eq:hatofdetrend}
\end{align}

Applying Eq.~\ref{eq:hatofdetrend} to the general equation of motion
Eq.~\ref{eq:general_eom_2} gives
\begin{equation}
  \partial_\tau \widehat{g_{detrend}}(\omega,\tau)
  +(1-\alpha)i\omega\widehat{g_{detrend}}
   = \widehat{f_{detrend}}.
\end{equation}
Multiplying by $\widehat{g_{detrend}}^*$ and taking the real part gives
\begin{equation}
  \thalf\partial_\tau \left|\widehat{g_{detrend}}(\omega, \tau)\right|^2
   = \Real\left[\widehat{g_{detrend}}^*\widehat{f_{detrend}}\right].
\end{equation}
Expanding $f(t)$ as a sum and using the linearity condition Eq.~\ref{eq:detrend_linear} gives the exact spectral budget:
\begin{equation}
  \thalf\partial_\tau \left|\widehat{g_{detrend}}(\omega, \tau)\right|^2
   = \sum_n \calA_n(\omega, \tau),\label{eq:spectral_budget_detrended}
\end{equation}
where
\begin{equation}
  \calA_n(\omega;\tau) \defeq  \Real\left[\widehat{g_{detrend}}^*\widehat{(A_n)_{detrend}}\right].
\end{equation}

Thus, we have shown that the temporal spectral budget remains balanced even when a detrending operation is incorporated.
While linear least squares was used in the previous derivation, any detrending operation that satisfies Eq.~\ref{eq:detofgdot} and Eq.~\ref{eq:detrend_linear} will also result in an exact equality in the spectral budget.
For example,
any linear filter acting upon windowed data works as well. That is, $h_{trend}$ can take the form of a convolution of the windowed signal with an impulse response function.  This can be expressed as $h_{trend}(t,\tau):=h(t) *_t(\sigma(t-\tau)f(t))$, where $*_t$ indicates convolution over the variable $t$, and where $h(t)$ is the impulse response function (i.e. inverse Laplace transform of the transfer function). Importantly, the taper function $\sigma(t-\tau)$ must again depend on $t$ and $\tau$ specifically through the combination $t-\tau$.

\section{Spectral kinetic energy transfers in fluid dynamics}\label{sec:fluids}

We now turn our attention to temporal and spatiotemporal spectral kinetic energy transfers in fluid dynamics applications. We derive for spatiotemporal and temporal transfers the nonlinear triad interactions already identified for spatial transfers in \citet{kraichnan67}. Additionally, we examine how flow features such as a mean flow affect the relationship between spatial, temporal, and spatiotemporal spectral transfers. We also derive the effects of isotropic sweeping on the temporal triad interactions,
particularly with regard to the locality of the triad interactions.
Many of these results will be applicable to dynamics in terms other than kinetic energy; we briefly discuss this in Appendix \ref{app:triadsofscalars}. 

Our starting point is the spatially Fourier transformed two- or three-dimensional Navier--Stokes equations defined by Eq.~\ref{eq:generaleom3dk}, although for the later discussion of triad interactions we need only assume the existence of a nonlinear advection term.
The spatiotemporal spectral equation of motion takes the form
\begin{align}
  \partial_\tau \mathcal{E}(\bk&,\omega,\tau; T) = \sum_n \calA_n(\bk,\omega,\tau; T) \\&= \calN(\bk,\omega,\tau; T) + \calD(\bk,\omega,\tau; T) + \calF(\bk,\omega,\tau; T), \label{eq:energy_budget_kw}
\end{align}
where
\begin{equation}
\mathcal{E}(\bk,\omega,\tau; T) =\thalf |\wht{\bu}(\bk,\omega,\tau; T)|^2
\label{eq:kineticenergydensity}
\end{equation}
is the kinetic energy density;
\begin{equation}
    \label{nonlinearadvectionspelledout}
  \calN(\bk,\omega,\tau; T) \defeq -\Real
  \left[
    \wht{\vb{u}}^*\cdot\left( \wht{\bnabla \Pi} + \wht{\bu\cdot\bnabla\bu}
    \right)
    \right]
\end{equation}
is the transfer due to nonlinear advection;
\begin{equation}
  \calD(\bk,\omega,\tau; T) \defeq
  -\Real
   \left[
    \sum_m \nu_m (-k^2)^m \mathcal{E}(\bk,\omega,\tau; T)
   \right]
\end{equation}
is the transfer due to dissipation;
and
\begin{equation}
  \calF(\bk,\omega,\tau; T) \defeq \Real
  \left[
    \wht{\bu}^*\cdot\wht{\boldf}
    \right]
\end{equation}
is the transfer due to forcing. 

\subsection{Doppler Shifting by a Mean Flow}\label{subsec:galileantheory}

\changes{Maybe repeat here what Eyink and Hussein had to say about the importance of Galilean invariance.}
We show how the spatiotemporal and temporal transfers change under a Galilean boost, which approximates the imposition of a mean flow.  Such a transformation is important because many flows of practical interest either exhibit a strong mean flow or are thought to exhibit isotropic sweeping, in which the smaller-scale structures are swept with minimal distortion by larger-scale structures. The following calculation best models an imposed mean flow when the external forcing is localized in $\bk$ space. In this case, the Galilean boost will affect only a small swath of $(\bk, \omega)$. The following calculation will not precisely model a mean flow when a forcing term sets a preferred frame of reference. For example, a forcing that is constant in time and doubly periodic in space in one frame would be distorted at all scales by a Galilean boost.

Consider two frames of reference in which velocity fields are $\bu^{(1)}$ and $\bu^{(2)}$, and that $\bV$ is the (constant) velocity of the second frame with respect to the first. 
The velocity fields and external forcing in the two frames are then related by  
\begin{align}
\bu^{(2)}(\bx,t)=\bu^{(1)}(\bx-\bV t,t)+\bV,\\
\boldf^{(2)}(\bx,t)=\boldf^{(1)}(\bx-\bV t,t).
\end{align} 
\changes{I might actually be able to improve the theory!  If the forcing does not change with $\bV$, then what?  Maybe it is not that hard to derive.}
As a means of setting notation, the energy budgets in frames $(1)$ and $(2)$ take the form 
\begin{align}
\frac{\partial}{\partial \tau}\mathcal{E}^{(m)}(\bk,\omega,\tau)  = \calN^{(m)}(\bk,\omega,\tau) &+ \calD^{(m)}(\bk,\omega,\tau) \nonumber\\&+ \calF^{(m)}(\bk,\omega,\tau),\label{eq:transfers_frames}
\end{align}
where $m=1,2$ labels the reference frame.

When $k\neq 0$, using the STFT with or without detrending, it can be shown that each of the terms in Eq.~\ref{eq:transfers_frames} transform according to
\begin{align}
\calA^{(2)}(\bk,\omega,\tau) & = \calA^{(1)}(\bk,\omega+\bV\bcdot\bk,\tau),\label{eq:transformation_rule} \; \text{ if } k\neq 0
\end{align}
where $\calA=\calN, \calD, \calF$, or $\partial_\tau \mathcal{E}$.
The transformation of each term is simply a $\bk$-dependent shift in the frequency variable.
The transfer in mode $(\bk,\omega)$ in frame (2) corresponds with the transfer in mode $(\bk,\omega+\bV\cdot\bk)$ in frame (1).


After integrating over all spectral angles in $\bk$-space
in order to obtain  $\calA(k,\omega,\tau)$, or after summing over all $\bk$ to obtain $\calA(\omega,\tau)$, we obtain the transformation rules
\begin{align}
\calA^{(2)}(k,\omega,\tau)  & := \int d\Omega_d\, k^{d-1} \calA^{(2)}(\bk,\omega,\tau)   \label{eq:transformation_rule_k_1} \\
& =  \int d\Omega_d\, k^{d-1} \calA^{(1)}(\bk,\omega+\bV\bcdot\bk,\tau;T), \label{eq:transformation_rule_k_2}  \\
\calA^{(2)}(\omega,\tau)  & := \sum_{\bk} \calA^{(2)}(\bk,\omega,\tau)  \\&=  \sum_{\bk} \calA^{(1)}(\bk,\omega+\bV\bcdot\bk,\tau)  \label{eq:transformation_rule_k_3}
\end{align}
where the integrals in Eq. \ref{eq:transformation_rule_k_1} and Eq. \ref{eq:transformation_rule_k_2} are over all angles in $d$-dimensional $\bk$-space.

According to Eq.~\ref{eq:transformation_rule_k_2} and Eq.~\ref{eq:transformation_rule_k_3}, after summing over multiple wavevectors $\bk$  the spectral content in mode $\omega$ in frame (2) corresponds to a sum of the spectral content over a range of wavevector-frequency modes in frame (1).
In general, the transfers $\calA^{(1)}(\bk,\omega,\tau)$ contribute substantially to the integral or sum only for wavenumbers $k<k_{max}$ for some wavenumber $k_{max}>0$. With $\omega$ fixed, the frequencies $\omega^\prime \defeq \omega + \bV\cdot\bk$ that contribute substantially to the sum will lie in the range $\omega - k_{max}V < \omega^\prime < \omega + k_{max}V$.
Thus, the effect of a mean flow is to ``spread'' out each frequency mode into nearby frequency--wavenumber modes. We assert that this result, represented specifically by Eq. \ref{eq:transformation_rule_k_2} and Eq. \ref{eq:transformation_rule_k_3}, is identical to the spectral broadening discussed by \citet{Tennekes75} and a generalization of the broadening shown by \citet{wilczek2012}, who implemented a Gaussian distributed sweeping velocity in addition to a mean velocity to show how $\bV\cdot\vb{k}$ acts as a Doppler shift for frequency spectra.

\subsection{Triad interactions}\label{subsec:propertiestheory}


Spatial triad interactions, $\mathcal{T}(\bk,\bp,\bq)$, give the rate of energy injection into wavevector mode $\bk$ due to nonlinear interaction with modes $\bp$ and $\bq$ \cite{kraichnan59, kraichnan66, Kraichnan71, kraichnan67}. The sum of all the triad interactions is the total transfer due to nonlinear advection. In this section, we define spatiotemporal and temporal triads interactions using asymmetric and symmetric diagrams defined with the STFT without detrending. Asymmetric diagrams were introduced in \cite{Dar2000} for plasma applications and independently reintroduced by \citet{skitka23} for wave turbulence in the ocean. In both cases, diagrams describe the individual energy exchanges between pairs of \textit{spatial} modes within a triad taken in isolation. Diagrams have an associated scalar value of energy transfer. In \citet{Dar2000} and \cite{Dar2001}, these diagrams describe the transfer of kinetic energy or magnetic energy in wavenumber space. In \citet{skitka23}, these diagrams are used solely to describe the transfer of kinetic energy in wavenumber.   Asymmetric diagrams describe a portion of the full interaction between three spectral modes; in particular, they define a direction of energy transfer between two distinct modes.
Instead of asymmetric diagrams, one can also decompose triad interactions into symmetric diagrams. Symmetric diagrams also have an associated scalar value of energy transfer. They describe the total energy injection into one mode from the two other modes in the triad interaction. We will use the notation $\mathcal{T}_u$ to represent asymmetric diagrams and $\mathcal{T}_s$ to represent symmetric diagrams. More explicitly, asymmetric spatiotemporal diagrams, $\mathcal{T}_u(\bk,\bp,\bq,\omega,\omega_{\bp},\omega_{\bq})$ give the rate of energy flow from wavevector-frequency mode $(\vb{q}, \omega_q)$ into wavevector-frequency mode $(\vb{k}, \omega)$, where $\vb{p} = \vb{k}-\vb{q}$ and $\omega_p = \omega - \omega_q$. Temporal diagrams $\mathcal{T}_u(\omega,\omega_{\bp},\omega_{\bq})$ give the rate of energy flow from frequency mode $\omega_{\bq}$ to $\omega$ with $\omega_p = \omega - \omega_q$. Symmetric spatiotemporal diagrams, $\mathcal{T}_s(\bk,\bp,\bq,\omega,\omega_{\bp},\omega_{\bq})$, are the rate of energy injection into wavevector-frequency mode $(\bk,\omega)$ due to nonlinear interaction with modes $(\bp,\omega_{\bp})$ and $(\bq,\omega_{\bq})$. Likewise, symmetric temporal diagrams, $\mathcal{T}_s(\omega, \omega_{\bp}, \omega_{\bq})$, give the rate of energy injection into frequency mode $\omega$ due to nonlinear interaction with modes $\omega_{\bp}$ and $\omega_{\bq}$. In this paper, we do not focus on possible uses of asymmetric versus symmetric diagrams. We instead use them to develop a potential method of applying tapers to the nonlinear advection term. In equations where both $\mathcal{T}_u$ or $\mathcal{T}_s$ can apply, we will leave off the subscript. We will show in the following section that symmetric spatiotemporal diagrams can be decomposed into the asymmetric spatiotemporal diagrams and symmetric temporal diagrams can be decomposed into asymmetric temporal diagrams. Furthermore, asymmetric and symmetric temporal diagrams can be derived from the asymmetric and symmetric spatiotemporal diagrams, respectively. 

If we use the STFT without a detrending operation, then we can apply the convolution theorem to show that the spatiotemporal spectral transfer due to the nonlinear advection term in Eq.~\ref{eq:energy_budget_kw} can be written as
\begin{widetext}
\begin{align}
\calN[\sigma]&(\bk,\omega,\tau;T) := \Real[\wht{\bu}^*[\sigma](\bk,\omega,\tau) \cdot (\wht{\bnabla \Pi}+\wht{\bu\cdot\bnabla\bu})[\sigma](\bk,\omega,\tau)]\label{eq:N}\\
&=\sum_{\bp,\bq} \int \: d\omega_{\bp}d\omega_{\bq} \mathcal{T}[\sigma_1, \sigma_2](\bk,\bp,\bq,\omega,\omega_{\bp},\omega_{\bq},\tau;T)\delta_{\bk-\bp-\bq,\bm{0}}\delta(\omega-\omega_{\bp}-\omega_{\bq}).
\end{align}
\end{widetext}
In this case, we choose to decompose the above equation in terms of asymmetric diagrams, setting $\mathcal{T}=\mathcal{T}_u$, defined explicitly by
\begin{widetext}
\begin{align}
\mathcal{T}_{u}[\sigma_1, \sigma_2](\bk,\bp,\bq,\omega,\omega_{\bp},\omega_{\bq},\tau)  :=  
\Bigg\{ 
    \begin{array}{cc}
        \:0, \: \: \:  \: \: \: \: \: \: \: \: \:  \: \: \: \: \: \:  \text{       if any of } \omega, \omega_{\vb{p}}, \omega_{\vb{q}}, \vb{k}, \vb{p}, \text{ or } \vb{q} = 0,\vspace{0.2cm}\\ \Imag\bigg[\Big(\widehat{\widetilde{\vb{u}}}^{*}[\sigma_1\sigma_2](\bk,\omega,\tau)\cdot \widehat{\widetilde{\vb{u}}}[\sigma_1](\bq,\omega_{\bq},\tau)\Big)\Big(\vb{k}\cdot\wht{\bu}[\sigma_2](\bp,\omega_{\bp},\tau) \Big)\bigg], \: \: & \text{otherwise}.
    \end{array}
\label{eq:unsym_triad_defined_check}  
    \end{align}
\end{widetext}
We cannot get the above results with a generalized detrending because of the convolution required to calculate Eq.~\ref{eq:N}. Specifically, calculating triad interactions with velocity fields that require detrending with functions that vary in time or space (even linear functions in time and space) introduces spurious terms to the triads calculated with Eq.~\ref{eq:unsym_triad_defined_check}. We show the inaccuracies of calculating the spectral transfers due to nonlinear advection with velocities detrended by linear functions in the right panel of Fig.~\ref{fig:result_detrending}. Constant offsets in space and time will not change nonlinear interactions. Eqn.~\ref{eq:unsym_triad_defined_check} does not admit self-interactions between modes with the same wavevector or frequency. 

In the above definitions, $\sigma_1$ and $\sigma_2$ are taper functions such that $\sigma_1\sigma_2 = \sigma$, where $\sigma$ is the taper function in Eq.~\ref{eq:N}. The constraint on taper functions emerges due to the convolution necessary to calculate the STFT of the nonlinear advection term. This introduces a conundrum unique only to the nonlinear advection term. The kinetic energy density and the transfers due to dissipation and forcing require two taper functions (i.e. in Eq. \ref{eq:kineticenergydensity}, each factor of velocity is tapered). There are no constraints on these taper functions and one can choose their taper of choice. The nonlinear advection term has three taper functions. One of the tapers is defined as the product of the other two, but it is not clear how to optimally choose the tapers $\sigma_1$ and $\sigma_2$. The condition $\sigma_1 \sigma_2 = \sigma$ implies that the only case where one can trivially set $\sigma_1 = \sigma_2$ without introducing a more extreme tapering is the case of applying a rectangular function as a taper. If ringing is a potential problem, the two tapers must be different and at least one taper must account for ringing. We suggest setting $\sigma_2 $ as the rectangular taper function (Eq.~\ref{eqn:recttaper}), which leaves $\sigma_1=\sigma$ such that $\sigma$ is the user's preferred taper function (i.e. We would use the Tukey function shown in Fig. \ref{fig: stftplot}). This amounts to only applying tapers to the two modes whose energy exchange is described by the diagram. This choice is physically appealing because it preserves the diagram's antisymmetry in the $(\vb{k}, \omega)$ and $(\vb{q}, \omega_{\vb{q}})$ modes. This antisymmetry, identified and discussed in Appendix A of \citet{skitka23} and Section IV of \citet{Dar2000}, allows asymmetric diagrams to be interpreted as an energy exchange between multiple spatial or temporal modes. With the tapering in Eq. \ref{eq:unsym_triad_defined_check}, one can compute such energy exchanges with minimal error caused by ringing. Additionally, this definition of tapered asymmetric diagrams allows us to define a tapered definition of symmetric wavenumber-frequency diagrams, $\mathcal{T} = \mathcal{T}_s$, that maintains their symmetry in $(\vb{p}, \omega_{\vb{p}})$ and $(\vb{q}, \omega_{\vb{q}})$. Written out explicitly, we would then define a symmetric wavenumber-frequency diagram as
\begin{widetext}
\begin{align}
\mathcal{T}_{s}[\sigma_1,\sigma_2](\bk,\bp,\bq,\omega,\omega_{\bp},\omega_{\bq},\tau)  :=\frac{1}{2} \Big[\mathcal{T}_{u}[\sigma_1, \sigma_2](\bk,\bp,\bq,\omega,\omega_{\bp},\omega_{\bq},\tau) + \mathcal{T}_{u}[\sigma_1, \sigma_2](\bk,\bq,\bp,&\omega,\omega_{\bq},\omega_{\bp},\tau)\Big].
\label{eq:triad_defined_check}  
\end{align}
\end{widetext}

 Eq.~\ref{eq:triad_defined_check}, despite being defined in terms of diagrams, is equivalent to the conventional definition of triadic interactions with the addition of tapered velocity fields \cite{kraichnan67, kraichnan59, Kraichnan71, Kraichnan80, waleffe92}. 

We leave $\tau$, $T$, and the choice of $\sigma$ out of the notation for Eq.~\ref{eqn:temporaltransfers}-\ref{eq:triad_galilean} because the forms of these equations is independent of the choice of these variables.


Using Parseval's theorem, we may obtain  $\mathcal{T}(\bk,\bp,\bq)$ from  $\mathcal{T}(\bk,\bp,\bq,\omega,\omega_{\bp},\omega_{\bq})$ by integrating over all frequencies  $(\omega$, $\omega_{\bp},$ $\omega_{\bq})$.
We may also obtain temporal diagrams by summing the spatiotemporal diagrams over all wavevectors:
\begin{align}
\mathcal{T}(\omega,\omega_{\bp},\omega_{\bq}) := \sum_{\bk,\bp,\bq}\mathcal{T}(\bk,\bp,\bq,\omega,\omega_{\bp},\omega_{\bq}).
\label{eqn:temporaltransfers}
\end{align}
In two spatial dimensions we can similarly derive diagrams for enstrophy, which are simply related to the energy diagrams:
\begin{align}
\label{eqn: enstrophytriad}
\mathcal{T}_{enstrophy}(\bk,\bp,\bq,\omega,\omega_{\bp},\omega_{\bq}) \defeq k^2\mathcal{T}(\bk,\bp,\bq,\omega,\omega_{\bp},\omega_{\bq}).
\end{align}
For two-dimensional turbulence, the spatial enstrophy transfers are straightforwardly related to the spatial energy transfers $\mathcal{T}(\bk,\bp,\bq)$:
\begin{align}
\label{eqn: enstrophytoenergy}
\mathcal{T}_{enstrophy}(\bk,\bp,\bq) & := \sum_{\omega,\omega_{\bp},\omega_{\bq}}k^2\mathcal{T}(\bk,\bp,\bq,\omega,\omega_{\bp},\omega_{\bq}) \\&= k^2\mathcal{T}(\bk,\bp,\bq).
\end{align}
However, the temporal enstrophy transfers
\begin{align}
\label{eqn: secondenstrophytoenergy}
\mathcal{T}_{enstrophy}(\omega,\omega_{\bp},\omega_{\bq})  := \sum_{\bk,\bp,\bq}k^2 \mathcal{T}(\bk,\bp,\bq,\omega,\omega_{\bp},\omega_{\bq})
\end{align}
have no clear relation to the temporal energy transfers $\mathcal{T}(\omega,\omega_{\bp},\omega_{\bq})$.
 This lack of a connection between $\mathcal{T}(\omega,\omega_{\bp},\omega_{\bq})$ and $\mathcal{T}_{enstrophy}(\omega,\omega_{\bp},\omega_{\bq})$ suggests that we cannot prove that  energy and enstrophy must be transferred in opposite directions 
 in $\omega$-space
 in the same way that one proves that energy and enstrophy must be transferred in opposite directions 
 in $k$-space
 for two-dimensional turbulence \cite{kraichnan67}.

 Symmetric temporal and spatiotemporal triads obey detailed conservation laws similar to those of spatial triads \cite{kraichnan67}. In two and three dimensions, written in terms of symmetric diagrams, detailed conservation of energy takes the form
 \begin{widetext}
\begin{align}
\mathcal{T}_s(\bk,\bp,\bq,\omega,\omega_{\bp},\omega_{\bq}) +\mathcal{T}_s(\bp,\bq,\bk,\omega_{\bp},\omega_{\bq},\omega) +\mathcal{T}_s(\bq,\bk,\bp,\omega_{\bq},\omega,\omega_{\bp}) & = 0,\\ 
\mathcal{T}_s(\omega,\omega_{\bp},\omega_{\bq}) +\mathcal{T}_s(\omega_{\bp},\omega_{\bq},\omega) +\mathcal{T}_s(\omega_{\bq},\omega,\omega_{\bp}) & = 0,
\end{align}
\end{widetext}
for spatial triads $\bk = \bp+\bq$ and temporal triads $\omega = \omega_{\bp} + \omega_{\bq}$.
Similarly, in two dimensions only, written with symmetric diagrams,  detailed conservation of enstrophy takes the form
\begin{widetext}
\begin{align}
k^2\mathcal{T}_s(\bk,\bp,\bq,\omega,\omega_{\bp},\omega_{\bq}) + p^2\mathcal{T}_s(\bp,\bq,\bk,\omega_{\bp},\omega_{\bq},\omega) +q^2\mathcal{T}_s(\bq,\bk,\bp,\omega_{\bq},\omega,\omega_{\bp}) & = 0,\\ 
\mathcal{T}_{s,enstrophy}(\omega,\omega_{\bp},\omega_{\bq}) +\mathcal{T}_{s,enstrophy}(\omega_{\bp},\omega_{\bq},\omega) +\mathcal{T}_{s,enstrophy}(\omega_{\bq},\omega,\omega_{\bp}) & = 0,
\end{align}
\end{widetext}
for symmetric spatiotemporal and temporal triads.
These rules imply that energy (and enstrophy if in two dimensions) is conserved within triads. 
\subsection{Locality of Triad interactions}\label{subsec:locality of triads theory}

In this section, we extend our discussion of mean flows and locality from Sec. \ref{subsec:galileantheory} to triad interactions. The effect of a mean flow, approximated again by a Galilean transformation, on the spatiotemporal diagrams is
\begin{widetext}
\begin{align}
\mathcal{T}^{(2)}(\bk,\bp,\bq,\omega,\omega_{\bp},\omega_{\bq}) = \mathcal{T}^{(1)}(\bk,\bp,\bq,\omega+\bV\cdot\bk,\omega_{\bp}+\bV\cdot\bp,\omega_{\bq}+\bV\cdot\bq),\label{eq:triad_galilean}
\end{align}
\end{widetext}
where $\bV$ is the velocity of reference frame (2) relative to frame (1).
The main implication of Eq.~\ref{eq:triad_galilean} is that the imposition of a mean flow, or sweeping, makes temporal triad interactions more nonlocal.  This should be expected as it has long been known~\citep{Tennekes75} that in the presence of sweeping, Kolmogorov-type arguments cannot be made for frequency spectra as they are made for wavenumber spectra, at least at sufficiently high wavenumber.  

The effect of isotropic sweeping on spectral transfers can be modeled by considering a single sweeping velocity magnitude $V = |\bV|$ and then integrating Eq.~\ref{eq:triad_galilean} over all directions of $\bV$.
General radially symmetric sweeping velocity distributions may then be handled by linear superposition.  

\begin{figure*}
    \includegraphics[width=0.8\textwidth]{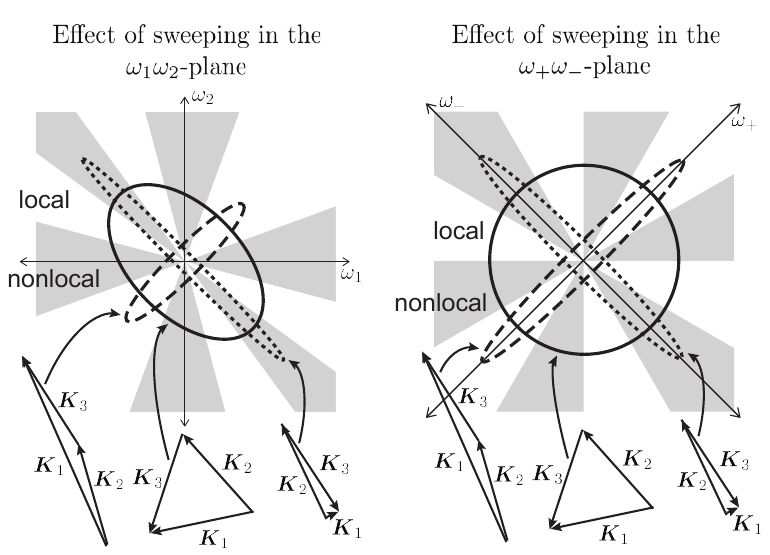}
    \caption{ The effect of isotropic sweeping on spatiotemporal triad interactions for three different spatial triads $(\bK_1,\bK_2,\bK_3)$.
  Sweeping redistributes a triad interaction at $(\Omega_1, \Omega_2, \Omega_3) = (0,0,0)$ onto an ellipse in the $\omega_+\omega_-$-plane, shown at the right. The projection of the ellipse in the $\omega_1\omega_2$-plane is shown at the left. Arrows indicate which spatial triad corresponds to which ellipse.
  Shaded regions indicate which temporal triads are nonlocal, as defined by Eq.~\ref{eq:local_def}.}\label{fig:ellipses}
\end{figure*}

  Suppose  that the transfer in the absence of sweeping (i.e. in frame 1) takes non-zero values only on the triad 
$(\bK_1,\bK_2,\bK_3,\Omega_{1},\Omega_{2},\Omega_{3})$, and let the sweeping velocity take the form
\begin{align}
 \bV = V\cos(\phi)\cos(\theta)\bxhat+ V\cos(\phi)\sin(\theta)\byhat + V\sin(\phi)\bzhat,
\end{align}
where $\phi$ is the angle between $\bV$ and the plane defined by the triad $(\bK_1,\bK_2,\bK_3)$, and $\theta$ is the angle between $\bK_1$ and the projection of $\bV$ onto that plane (making $\bxhat = \bK_1 / |\bK_1|$). For consistency, we re-index the frequencies in frame 2 with numbers, as in ($\omega_1$, $\omega_2$, $\omega_3$).
Restricting the analysis to two spatial dimensions, so that $\phi=0$,  the effect of integrating Eq.~\ref{eq:triad_galilean} over all $\theta$ (i.e.~the effect of isotropic sweeping by a single $V$) is to redistribute the transfer at $(\Omega_1,\Omega_2,\Omega_3)$ onto an ellipse in $(\omega_{1},\omega_{2},\omega_{3})$-space centered at $(\Omega_1,\Omega_2,\Omega_3)$, with the ellipse lying in the plane defined by $\omega_{\bk}=\omega_{\bp}+\omega_{\bq}$.  In that plane, one may define an orthonormal coordinate system with components $(\omega_+,\omega_-)$ given by $\omega_+=\sqrt{3/2}(\omega_{2}+\omega_{3})=\sqrt{3/2}\omega_{1}$ and $\omega_-=\sqrt{1/2}(\omega_{2}-\omega_{3})$.

Consider triads that satisfy $K_2=K_3$, where by definition $K_n := |\bK_n|$.  The more general case with arbitrary $\bK_2$ and $\bK_3$
is not substantially more illuminating. Depending on the angle between $\bK_2$ and $\bK_3$, the spatial triad $(\bK_1,\bK_2,\bK_3)$ can be relatively local or nonlocal, with maximal locality obtained when the wavevectors form an equilateral triangle ($K_1=K_2=K_3$).

The resulting ellipse in the $\omega_+\omega_-$-plane will take the form
\begin{equation}
\frac{(\omega_+ - \Omega_+)^2}{(K_+V)^2} + \frac{(\omega_- - \Omega_-)^2}{(K_-V)^2} = 1, \label{eq:special_ellipse_equation}
\end{equation}
where $\Omega_+ := \sqrt{3/2}(\Omega_2 + \Omega_3)$ and $\Omega_- := \sqrt{1/2}(\Omega_2 - \Omega_3)$ and, similarly, $\bK_+ := \sqrt{3/2}(\bK_2 + \bK_3)$ and $\bK_- := \sqrt{1/2}(\bK_2 - \bK_3)$.
The axes of the ellipse will be aligned with the $\omega_+$- and $\omega_-$-axes, and the corresponding major/minor axes will be $K_+V$ and $K_-V$.
In terms of $\bK_1$ and $\bK_2$, the major/minor axes can be written as
\begin{eqnarray}
  K_+V &=& \sqrt{6}K_3|\cos(\theta_{23}/2)|V, \label{eq:major_axis} \\
  K_-V &=& \sqrt{2}K_3|\sin(\theta_{23}/2)|V, \label{eq:minor_axis}
\end{eqnarray}
where $\theta_{23}$ is the angle between $\bK_2$ and $\bK_3$.

We use the locality measure originally defined by \citet{lesieur1997} to quantify locality of triads; any temporal triad that satisfies 
  \begin{equation}
\gamma := \frac{\max(|\omega_{1}|,|\omega_{2}|,|\omega_{3}|)}{\min(|\omega_{1}|,|\omega_{2}|,|\omega_{3}|)}<3\label{eq:local_def}
  \end{equation}
is defined as local. 
Temporal triads are inherently less local than spatial triads due to the fact that frequencies are scalars.  While it is possible to simultaneously satisfy $|\bk|$=$|\bp|$=$|\bq|$ and  $\bk=\bp+\bq$, it is not possible to satisfy both $|\omega_{1}|=|\omega_{2}|=|\omega_{3}|$ and $\omega_{1}=\omega_{2}+\omega_{3}$.
Thus, the smallest possible value of $\gamma$ for temporal triads is $\gamma = 2$ rather than $\gamma = 1$.
Our choice of threshold for nonlocality at $\gamma=3$ is natural because it divides the $\omega_+\omega_-$-plane into equal areas (Fig.~\ref{fig:ellipses}).

Fig.~\ref{fig:ellipses} demonstrates the effect of isotropic sweeping on three different spatiotemporal triads with $(\Omega_1, \Omega_2, \Omega_3) = (0,0,0)$.  The spatial triads for the three cases are $K_1=K_2=K_3$ (equilateral triangle), $\bK_2\approx\bK_3$, and $\bK_2\approx -\bK_3$.
The figure shows the swept transfers in the  $\omega_+\omega_-$-plane as well as the projection of the swept transfers onto the $\omega_{\bp}\omega_{\bq}$-plane.
Shaded regions correspond to nonlocal temporal triads, while unshaded regions correspond to local temporal triads, according to Eq.~\ref{eq:local_def}.

According to Eq.~\ref{eq:major_axis} and Eq. \ref{eq:minor_axis}, the major and minor axes will be equal (\ie the ellipse will be a circle) precisely when $\theta_{23} = \pm 2\pi/3$. In other words, restricted to the case $K_2=K_3$, the ellipse will be a circle if and only if the triad $(\bK_1, \bK_2, \bK_3)$ forms an equilateral triangle. In contrast, the minor axis will be zero whenever $\theta_{23} = 0$ or $\theta_{23}=\pi$. When $\theta_{23} = 0$ (\ie $\bK_2$ is parallel to $\bK_3$) the major axis will lie along the $\omega_+$-direction, which corresponds to local frequency triads. When $\theta_{23} = \pi$ (\ie $\bK_2$ and $\bK_3$ are antiparallel) the major axis will lie along the $\omega_-$-direction, which corresponds to nonlocal frequency triads.


Moreover, according to Eq.~\ref{eq:major_axis} and Eq. \ref{eq:minor_axis}, the major/minor axes of the ellipses in the  $\omega_+\omega_-$-plane are directly proportional to the sweeping velocity $V$ and the size $K$  of the spatial triad.  Thus, for sufficiently large $V$ or $K$ the ellipse will be large enough to encompass many regions of locality and nonlocality in the $\omega_+\omega_-$-plane, causing any originally local temporal triad to be distributed into both local and nonlocal triads.

This simple example can be generalized to the case where $\Omega$'s are non-zero if they are still approximately equivalent such that $\gamma \approx 1$ when defined on $\Omega$'s. If $\Omega$'s are non-zero, we can recast the coordinate system of Fig.~\ref{fig:ellipses} as $(\omega_+ - \Omega_+,\omega_- - \Omega_-)$. This change in coordinates would be accompanied subtracting $\Omega_1$, $\Omega_2$, and $\Omega_3$ from $\omega_1$, $\omega_2$, and $\omega_3$ respectively. If the $\Omega$'s are nonlocal, shifting the $\omega_+\omega-$ coordinates system could change the regions of locality and nonlocality relative to the origin of such a coordinate plane. This would be the case where isotropic sweeping localizes nonlocal spatiotemporal triads, as opposed to delocalizing local spatiotemporal triads as shown in Fig.~\ref{fig:ellipses}.

The three-dimensional case is a simple extension of the two-dimensional case discussed above. In three dimensions the effect of integrating over all $\theta$ is to form an ellipse in the  $\omega_+\omega_-$-plane  exactly as in the two-dimensional case, except that $V$ is replaced by $V\cos(\phi)$.  So, upon integrating over all $\phi$, the interior of the circle is filled as well, albeit non-uniformly with lowest density near the center of the circle.

\section{Numerical investigation: 2D turbulence}\label{sec:numerics}


In this section, we numerically simulate the two-dimensional Navier--Stokes equations with a general dissipation and a forcing term.
We use the spatiotemporal spectral transfer diagnostic developed in Sec. \ref{subsec:spatialtemporaltheory} to investigate the dynamics of energy and enstrophy in statistically equilibrated simulations forced at four different frequencies.  To test the robustness of this diagnostic, we also show the effects of varying the window size and sampling rate on the numerically calculated spectral transfers, and we demonstrate the effect of detrending when the simulation output is non-stationary.
\subsection{Numerical setup}\label{sec:numerical_setup}




In spectral space, assuming spatially periodic boundary conditions, the simulated equation takes the following form:
\begin{equation}
\frac{\partial}{\partial t} k^2 \widetilde{\psi}(\bk,t) = \widetilde{J(\psi,\nabla^2\psi)}(\bk,t) - \widetilde{D}(\bk,t) -\widetilde{F}(\bk,t),\label{eq:num_eom}
\end{equation}
where the forcing term $\widetilde{F}$ and the dissipation term $\widetilde{D}$ will be described below.  The width of the square, periodic domain is $L$.

\begin{figure*}
\begin{center}
    {\includegraphics[width=\textwidth]{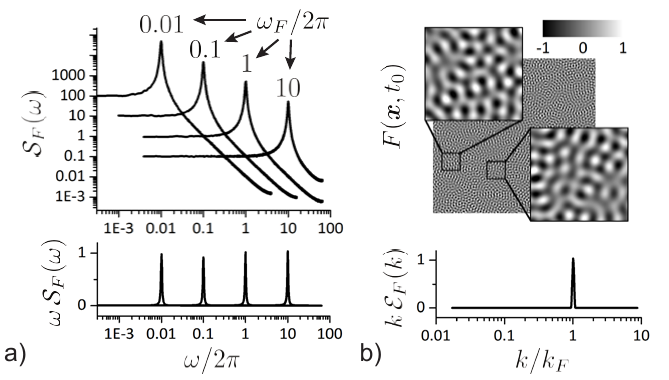}}
\end{center}
  \caption[Properties of the forcing, which is narrowband in both frequency and wavenumber.]{Properties of the forcing, which is narrowband in both frequency and wavenumber: (a)~Temporal spectral densities $\mathcal{S}_F(\omega)$ of the four different forcing frequencies (top) and  area-preserving spectral densities  $\omega \mathcal{S}_F(\omega)$ (bottom). (b)~Snapshot of the forcing $F(\bx,t_0)$ at initial time (top) and the corresponding area-preserving spatial spectral density $k\mathcal{E}_F(k)$  (bottom), which was the same for all four forcing frequencies. The scales of the vertical axes are arbitrary.  }
  \label{fig:forcing}
\end{figure*}


The forcing term $\widetilde{F}(\bk,t)$ is narrowband in both wavenumber and frequency. The forcing is statistically spatially isotropic with peak spectral amplitude at the forcing wavenumber $k_F$ and is statistically stationary with peak spectral amplitude at the forcing frequency $\omega_F$.
Fig.~\ref{fig:forcing} shows the temporal and spatial spectral densities of the forcing for four values of $\omega_F$.
Fig.~\ref{fig:forcing} also shows a snapshot of the forcing in physical space for comparison with snapshots of the stream function and vorticity shown in Fig.~\ref{fig:result_snapshots}.
A full description of the forcing is provided in Appendix~\ref{app:forcing}.

\changes{the following paragraph was cut and pasted from the old theory section, and I removed it entirely:
Regarding the generalized viscosity, it should be noted that $n=1$ corresponds to regular viscosity, $n=0$ corresponds to Ekman drag (or Rayleigh friction, or Hartmann friction, or air friction), and $n<=0$ and $n>1$ corresponds to what we will call \emph{inverse viscosity} and \emph{hyper-viscosity}, respectively.  We reserve the term \emph{hypo-viscosity} for the case $0<n<1$, while noting that this terminology is not used consistently in the literature.  If one chooses to employ an inverse viscosity ($n<-1$), then one must restrict the space of solutions so that the Laplacian ($\nabla^2$) has an inverse.  For example, in the case of periodic boundary conditions one simply needs to require that the $\bk=0$ Fourier mode of $\bu(\bx,t)$ or $\psi(\bx,t)$ is zero.}

For dissipation at the small and large scales, we apply two cutoff wavenumber filters similar in form to hyperviscosity and inverse viscosity. Specifically, we set
\begin{widetext}
\begin{equation}
\widetilde{D}(\bk,t) := \nu_{6}(k-k_6)^{6}H(k-k_6) k^2\widetilde{\psi}(\bk,t) + \nu_{-6}(k^{-1}-k_{-6}^{-1})^{6}H(k_{-6}-k) k^2\widetilde{\psi}(\bk,t),\label{eq:dissipation}
\end{equation}
\end{widetext}
where $H(\cdot)$ is the Heaviside step function.
The first term in Eq.~\ref{eq:dissipation} can be thought of as a hyperviscosity applied to wavenumbers $k>k_6$, while the second term in can be thought of as an inverse viscosity applied to wavenumbers $k<k_{-6}$. This choice ensures a wide inertial range between $k_{-6} < k < k_6$.  
We set $\nu_{6} = 1.3\times 10^{-11}$, $k_{6}= 366\approx 0.715 k_N$, $\nu_{-6} =1.2\times 10^{9}$ , and $k_{-6}=7.9\approx 0.0154 k_N$, where $k_N$ is the Nyquist wavenumber.
The cut-off wavenumber $k_{-6}$ for the  inverse wavenumber filter was set sufficiently high so that the stream function would not be significantly spatially correlated with itself halfway across the domain. 
 This ensures that the lowest frequencies observed in the system are not appreciably affected by the spatial periodicity. \changes{double check this}

To simulate Eq.~\ref{eq:num_eom} we use a parallelized pseudo-spectral method ~\citep{Canuto2007a,Canuto2007b}, with domain size $L=2\pi$ and spatial resolution of $1024^2$.  
The vorticity is iterated using third-order Adams--Bashforth time-stepping, while the forcing term is iterated separately using Euler's method with a smaller time step.  While technically the inclusion of a stochastic forcing term changes the strong (root-mean-square) convergence of the numerical scheme from third order to first order, the error is better represented by $O(\Delta t^3 + \Delta t/M)$, where $M$ is the ratio of the vorticity-term time step to the forcing-term time step.  This effectively third-order numerical method  (for sufficiently large $M$) resembles numerical methods developed for stochastic systems with small noise ~\citep{Buckwar06,Buckwar07}.   We found that $M=1$ was sufficient for the convergence of all statistics we present.  
The time-stepping increment was either  $\Delta t = 2^{-12}$ or $2^{-13}$ for all runs.

Most runs were not fully de-aliased, but the wavenumber filter 
removed practically all enstrophy at wavenumbers larger than $5/6$ times the Nyquist wavenumber.
 A single fully de-aliased simulation was conducted for comparison, and there was no significant change in the results.

 



We simulated the Eq.~\ref{eq:num_eom} for four different choices of the forcing frequency, $\omega_F/2\pi\in\{0.01,0.1,1,10\}$, with corresponding  integral time scales  of $\tau_F=5\times 2\pi/\omega_F$.
For each of the four forcing frequencies, we created three time series (each approximately one terabyte in size) by sampling the simulation at three different sampling rates.  Each time series consisted of $32,768$ snapshots, and the three time series durations were $t_{max}\in\{ 256, 1024, 4096 \}$.
 For each time series the window size used in calculating the spectral transfers was equal to the entire length of the time series.  Thus, the calculated spectral transfers took the form $\calA(\bk,\omega,t_{max}/2;t_{max})$, where the central time was set equal to the center of the data set, $\tau = t_{max}/2$.


\begin{figure*}
\includegraphics[width=0.9\textwidth]{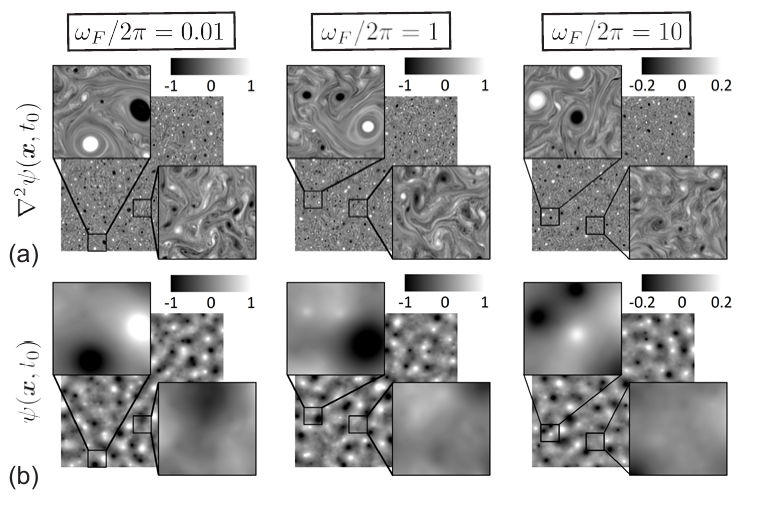}
\caption[Stream function and vorticity snapshots.]  { Stream function and vorticity snapshots: (a) Snapshots of  vorticity $\nabla^2\psi(\bx,t_0)$ after statistical equilibrium is reached for three of the four forcing frequencies. Magnified insets show regions with and without strong vortices.  The vorticity snapshot for the missing forcing frequency ($\omega_F/2\pi=0.1$)  is visually and quantitatively similar to the case $\omega_F/2\pi=1$. Only relative scale is important, so each plot uses the same (but otherwise arbitrary) units. Note the decreased vorticity range in the rightmost plots  ($\omega_F/2\pi=10$). (b) Snapshots of stream function $\psi(\bx,t_0)$ with corresponding insets.  
}
\label{fig:result_snapshots}
\end{figure*}

\begin{figure}
    \includegraphics[width=\columnwidth]{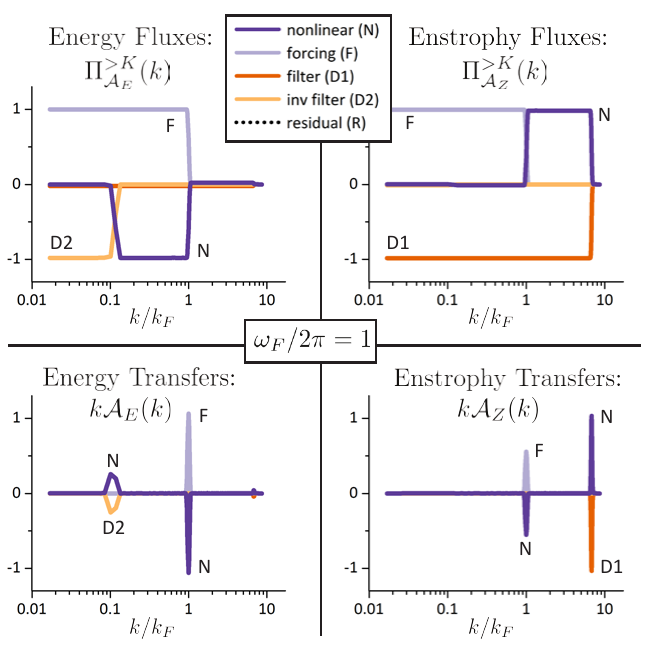}
  \caption{
    Spatial spectral fluxes of energy and enstrophy (top) and the corresponding spatial spectral transfers (bottom) for the $\omega_F/2\pi=1$ simulation. A positive transfer at wavenumber $k$ corresponds to the injection of energy (or enstrophy) at that wavenumber. A positive flux corresponds to a downscale transfer 
    to larger $k$, and a negative flux corresponds to an upscale transfer 
    to smaller $k$. A negatively sloping flux corresponds to a positive transfer, while a positively sloping flux corresponds to a negative transfer. The statistical noise in the fluxes and transfers is very small by design. The noise in the forcing is quite weak and the range of forcing and dissipation ranges are quite small. Furthermore, they have been averaged over $32,768$ snapshots.   
}
\label{fig:result_fourforcings_k}
\end{figure}

\subsection{Spatial transfers and fluxes}

In this section, we calculate the spectral transfers and fluxes. The spectral fluxes capture all the contributions to the kinetic energy evolution at all wavenumbers or frequencies above a threshold. Historically, this quantity has been defined in wavenumber space either as \citep{ScottArbic2007}
\begin{equation}
  \Pi_{\calA_n}^{>}(k,t) := \int_k^\infty dk^\prime \; \calA_n(k^\prime,t),\label{eq:flux_kt_def}
\end{equation}
or by the time average of the above equation,
\begin{equation}
\Pi_{\calA_n}^{>}(k) := \int_k^\infty dk^\prime \; \calA_n(k^\prime), \label{eq:flux_k_def}
\end{equation}
where $\calA_n(k)$ is the time-average of $\calA_n(k,t)$.
Alternatively, with the development of temporal spectral transfers, we can define a temporal spectral flux:
\begin{equation}
    \label{eq:flux_omega_def}
  \Pi_{\calA_n}^{>}(\omega,\tau;T) := \int_\omega^\infty d\omega^\prime \; \calA_n(\omega^\prime,\tau; T).
\end{equation}

While the theory presented earlier in this paper focuses primarily on spectral transfers, we show both transfers and fluxes because some readers will be more familiar with spectral fluxes. Spectral fluxes have the same frequency range as spectral transfers, determined by $T$ and sampling rate as with spectral transfers. Because the value of spatiotemporal transfers at different frequencies can vary due to the width of frequency bins (as mentioned in Sec. \ref{subsec:spatialtemporaltheory}), the value of spatiotemporal fluxes will also vary due to the width of frequency bins.  Fig.~\ref{fig:result_fourforcings_k} shows the time-averaged spatial spectral transfers and fluxes for the case $\omega_F/2\pi=1$.
The spatial spectral transfers and fluxes for the other three forcing frequencies are qualitatively similar, with the main difference being that the run forced at the highest frequency exhibited transfers and fluxes of smaller magnitude. 
Some readers may be accustomed to significantly noisier spatial spectral transfers. Our transfers are relatively smooth because we average over 32,768 snapshots in time. 

In the bottom left plot of Fig.~\ref{fig:result_fourforcings_k}, showing energy transfers, we see the transfer by forcing (lavender-grey curve, labelled ``F") balanced out by a corresponding negative transfer by nonlinear advection (dark purple curve, labelled ``N") at the forcing wavenumber. We see a transfer of this energy to a lower wavenumber, where the inverse wavenumber filter (yellow curve, labelled ``D2" on the plot and labelled ``inv filter" in the legend) has a negative transfer.  The upper left plot, showing the corresponding energy fluxes reflects the integrals of the transfer curves. The negative values of the nonlinear advection curve reflects a transfer from high wavenumber to low wavenumber. The forcing curve is positive since it represents an injection of energy and the inverse filter is negative because it represents the dissipation of energy from the system. The bottom right plot shows enstrophy transfers, where we see similar behavior at the forcing wavenumber. However, enstrophy is injected at a \textit{higher} wavenumber by nonlinear advection. There, the enstrophy is dissipated by the wavenumber filter (orange curve, labelled ``D1" on the plot and ``filter" in the legend). This transfer at higher wavenumber is captured in the top right plot of Fig.~\ref{fig:result_fourforcings_k} by the positive nonlinear advection curve. These results are expected for two-dimensional turbulence. Energy is injected at a forcing wavenumber and is dissipated at small wavenumber (large spatial scales). Enstrophy, on the other hand, is injected at a forcing wavenumber and is dissipated at large wavenumber (small spatial scales). 

\subsection{Temporal spectral fluxes}

We study temporal spectral fluxes of energy and enstrophy for all four forcing frequencies in Fig.~\ref{fig:result_fourforcings_w}. We focus on spectral fluxes because the corresponding spectral transfers are significantly noisier and potentially more difficult to interpret. To illustrate this point, we also show the temporal spectral transfers of energy for the highest two forcing frequencies in Fig.~\ref{fig:result_fourforcings_w}.

Fluxes are shown for each of three sampling rates, indicated by different line styles. Solid lines correspond to the largest sampling rate , dashed lines correspond to the intermediate sampling rate, and dotted lines correspond to the lowest sampling rate. Sampling rates are defined by a fixed number of snapshots (32,678) taken over a time duration $t_{max}$; thus, the largest sampling rate corresponds to the smallest $t_{max}$ and vice-versa. There is close agreement between the results with the different sampling rates, except at the highest and lowest frequencies resolved by each time series.  The close agreement indicates that the temporal spectral transfers and fluxes are reasonably robust diagnostics for time series that resolve most but not all dynamical time scales.

In contrast with the spatial spectral fluxes, the temporal spectral fluxes differ considerably between simulations with different forcing frequencies. Just as with wavenumber fluxes, we use the nonlinear advection term's (dark purple curves, labelled ``N") fluxes as an indicator of the overall flow of energy and enstrophy for the fluid. For temporal fluxes, this term reflects the flow of a quantity from time scale to time scale \cite{Kraichnan71, skitka23}. At the lowest forcing frequency, $\omega_F/2\pi=0.01$, the energy flux is positive, indicating a net transfer of energy to frequencies higher than the forcing frequency. When $\omega_F/2\pi=1$, the energy flux is negative, indicating a net transfer of energy to frequencies below the forcing frequency. In the intermediate case, when $\omega_F/2\pi=0.1$, there are both upscale and downscale transfers of energy. In all three of these cases, the energy is injected at the forcing frequency and then transferred to the range of frequencies over which large-scale dissipation occurs. We observe this in the the energy and enstrophy flux plots for the three lowest forcing frequencies, where the flux by forcing (lavender grey plot, labelled ``F"), sharply decreases at the forcing frequency. In the energy transfer plot (left most column, top plot), this is reflected in the sharp peak in the transfers by forcing. 
At the highest forcing frequency, $\omega_F/2\pi=10$, energy and enstrophy are not injected at the forcing frequency. In the energy transfer plot at the bottom left, we see both positive and negative peaks in the transfers by forcing and then a surge of forcing amplitude at a range of lower frequencies. In the middle plot of the bottom row, we observe that the energy flux by forcing decreases to zero at frequencies almost an order of magnitude lower than the forcing frequency. The same behavior is visible in the enstrophy flux, in the bottom right plot. We interpret this as the forcing frequency being too high for the fluid to respond at the forcing frequency. Instead, energy and enstrophy are injected across a wider range of frequencies below the forcing frequency. From there, energy is transferred to relatively lower frequencies while enstrophy is transferred to relatively higher frequencies. 
As with fluxes and transfers in wavenumber, the wavenumber filter is responsible for dissipating the enstrophy (here, at high \textit{frequencies}) while the inverse wavenumber filter is responsible for dissipating the energy (here, at frequencies higher or lower than the forcing frequency). 
Our results show that while the direction of energy flux varies depending on the frequency of the forcing, enstrophy appears to solely move to higher frequencies. In other words, the dissipation of enstrophy always occurs at frequencies above the injection frequency, which is evidenced by the positive nonlinear flux for all cases. In the simulation with the highest forcing frequency, the fluid cannot respond to the dominant time-scale of the forcing.  Turbulence still develops but at a less energetic level, as was previously found in other studies of ``modulated" turbulence in three-dimensions ~\citep{Lohse00,Hooghoudt01,Heydt03,Cadot03,Kuczaj06,Kuczaj08}.  Because the fluid cannot respond to arbitrarily high frequencies (without changing the forcing amplitude), there is an upper limit for the frequency at which enstrophy can be injected.  If this limiting injection frequency is lower than the highest frequencies associated with the enstrophy dissipation range, then there cannot be a transfer of enstrophy to frequencies below the injection frequency. Thus, while energy can be transferred either to smaller or larger timescales relative to the period of the forcing, enstrophy appears to only move to small timescales in two-dimensional turbulence.
\subsection{Temporal spectral transfers}
\begin{figure*}
\includegraphics[width=0.8\textwidth]{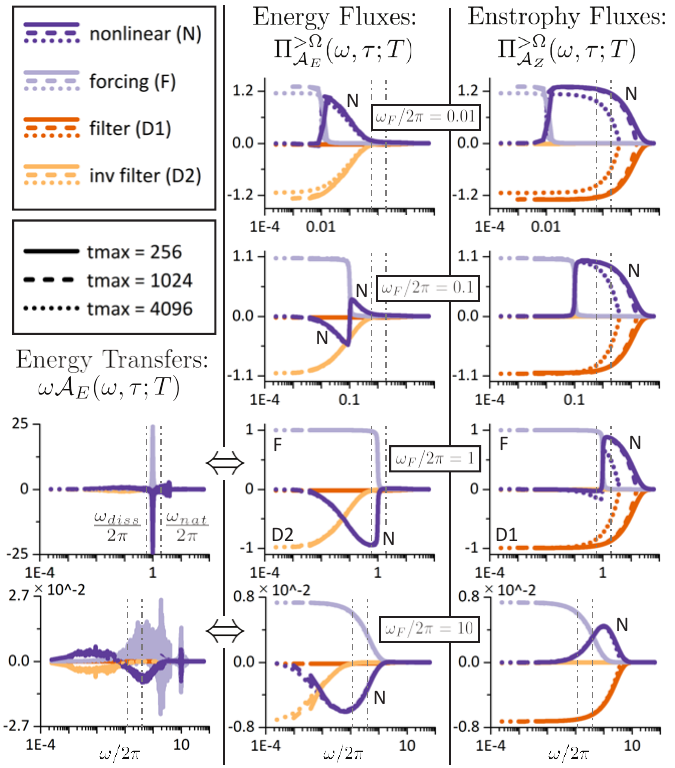}
\caption{ Temporal spectral fluxes of energy and enstrophy for each of the four forcing frequencies, and temporal spectral transfers of energy for the two highest forcing frequencies.
Each plot contains three curves, one solid, one dashed, and one dotted. The curves vary by duration $t_{max}$, with longer durations using lower sampling rates. The natural frequency $\omega_{nat}$ and external frequency $\omega_{ext}$ are indicated by dashed lines. For each of the four forcing frequencies (rows) temporal spectral \emph{fluxes} of energy (middle column) and enstrophy (right column).  For the two highest forcing frequencies, the temporal spectral \emph{transfers} are shown as well (left column), corresponding to the spectral fluxes directly to their right.  The upper legend defines each term in the spectral budget.  The lower legend indicates values of $t_{max}$ and thus the simulation sampling rate. The two frequencies derived \emph{post-priori} (see Appendix \ref{app:nondim}), $\omega_{diss}/2\pi$ and $\omega_{nat}/2\pi$ are indicated by gray vertical dash-dotted lines. For all plots, $\omega_{diss}/2\pi < \omega_{nat}/2\pi$.}
\label{fig:result_fourforcings_w}
\end{figure*}

\begin{figure}
    \includegraphics[width=0.9\columnwidth]{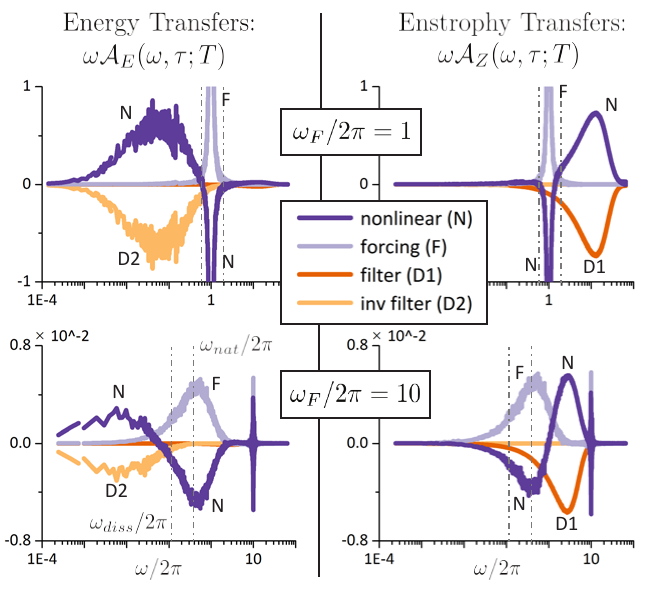}
    \caption{
  Temporal spectral transfers of energy and enstrophy for the simulations forced at the two highest forcing frequencies, $\omega_F/2\pi=1$ and $\omega_F/2\pi=10$.
  Each plot is created using the same three sampling rates as in Fig.~\ref{fig:result_fourforcings_w}, but the noisiest regions of each curve have been removed.
  In the $\omega_F/2\pi=10$ simulation, the transfers near $\omega/2\pi=10$ are noisy but integrate roughly to zero, which is proved by the smoothness of the fluxes near $\omega/2\pi=10$. The two frequencies derived \emph{post-priori} (see Appendix \ref{app:nondim}), $\omega_{diss}/2\pi$ and $\omega_{nat}/2\pi$ are indicated by gray vertical dash-dotted lines. The gaps in the bottom left plot are due to differently sampled curves that do not overlap at that frequency. The curve with $t_{max} = 1024$ ends before the gap and the dashed plots with longer $t_{max} = 4096$ continues to lower frequencies.}
\label{fig:result_fourforcings_transfers_w}
\end{figure}

In  Fig.~\ref{fig:result_fourforcings_transfers_w}, we show the temporal spectral transfers corresponding to the fluxes in Fig.~\ref{fig:result_fourforcings_w}.   We reduce the range of the y-axis, which corresponds to the scale of the energy or enstrophy, to display fine-scale features of each transfer.

We use the longest duration dataset for the lowest frequencies, the shortest duration dataset for the highest frequencies, and the intermediate duration dataset for the intermediate frequencies.  This choice removes the noise due to aliasing at the highest frequencies and removes incorrect transfers at the lowest frequencies.
When the fluid is forced at $\omega_F/2\pi=1$, energy and enstrophy are injected within a narrow range around the forcing frequency.
When the fluid is forced at $\omega_F/2\pi=10$, there appear to be both positive and negative transfers at the forcing frequency, but these average roughly to zero as demonstrated by the nearly zero flux at the same frequency in Fig.~\ref{fig:result_fourforcings_w}. Instead, both energy and enstrophy are injected over a relatively wide range of frequencies centered at approximately $\omega/2\pi=0.4$.

Broadly speaking, these temporal spectral transfer plots lend insight into whether energy is being injected or removed from different timescales. In both plots in the left column of Fig.~\ref{fig:result_fourforcings_transfers_w}, we can examine the nonlinear advection term (dark purple curve labelled ``N") and see that energy is removed from the forcing frequencies and then injected at lower frequencies. For the two plots in the right column of Fig.~\ref{fig:result_fourforcings_transfers_w}, we see that enstrophy is removed from forcing frequencies and then injected at higher frequencies. 

Similar plots have been calculated in other papers from our research group. Figure 14a in \citet{Arbic14} shows temporal spectral transfers calculated in a two-layer quasigeostrophic turbulence simulation. This plot shows similar behavior to the nonlinear advection term in Fig.~\ref{fig:result_fourforcings_transfers_w}; in the top-layer of the simulation, kinetic energy is removed from high frequencies and deposited over a range of lower frequencies. 

In general, temporal spectral transfer plots are useful for showing how different phenomena are responsible for injecting or removing energy at different frequencies. This can be especially useful in studying simulations where energy or other budgets have many different components contributing at different time scales. For example, Figure 14a in \citet{Arbic14} shows the temporal spectral transfers due to nonlinear advection, forcing, friction, and available potential energy. Such plots could especially be useful in coupled ocean-atmosphere climate simulations, where different phenomena could contribute to energy or temperature variance budgets at scales ranging from daily to decadal. 

\subsection{spatiotemporal spectral transfers}
\begin{figure}
\includegraphics[width=0.85\columnwidth]{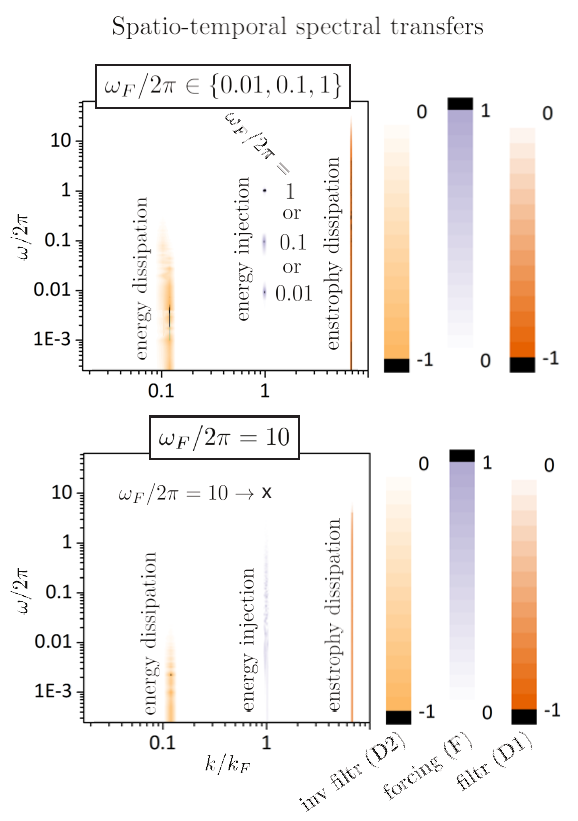}
\caption{Spatiotemporal spectral transfers for the three lowest forcing frequencies (top) and the highest forcing frequency (bottom).  To reduce statistical noise, the lower half of each plot is based on the simulation output with the longest duration($t_{max}=4096$), while the upper half of each plot is based on the simulation output with the shortest duration ($t_{max}=256$). Transfers of energy and enstrophy do not substantially overlap in $(k,\omega)$-space except at the forcing frequencies, where we plot energy transfers only.  Dissipation transfers were quantitatively similar for the three lowest forcing frequencies, so we simply show the result for $\omega_F/2\pi=1$ in the upper plot.  In the bottom plot, the symbol ``x'' marks the location of $(k_F,\omega_F)$, which is notably outside the range of frequencies in which energy (and enstrophy) is injected by the forcing. The nonlinear term (not shown) is approximately the negative of the sum of the other terms.}
\label{fig:result_fourforcings_kw}
\end{figure}

Fig.~\ref{fig:result_fourforcings_kw} shows the full spatiotemporal spectral transfers for all four forcing frequencies.  The top plot combines the results for the three lowest forcing frequencies. The combination of these three cases is possible because  the transfers due to dissipation remain largely unchanged while the transfers due to the forcing do not overlap.  The transfers due to the nonlinear term are not shown, because they can be calculated using the fact that the sum of all the transfers is zero everywhere in $(k,\omega)$-space (the residual is effectively zero).
The vertical smearing at fixed wavenumber in the bottom plot of Fig.~\ref{fig:result_fourforcings_kw} again shows that energy injection takes place at a range of frequencies well below the intended forcing frequency.

We show three colorbars, each for different sources of transfers. We have an light orange colorbar for the inverse wavenumber filter that ranges from $-1$ to $0$, a lavender grey colorbar for the forcing that ranges from $0$ to $1$, and an darker orange colorbar for the wavenumber filter ranging from $0$ to $1$. All of the colorbars are white where their corresponding field is $0$ and black where their corresponding field has a magnitude (absolute value) of $1$ or higher. We see black spots in the figure where that field is valued at $1$ or significantly higher. These are especially clear in the top and bottom plots in at $0.1 k_F$, where we observe the light orange streak denoting the inverse wavenumber filter (labelled ``energy dissipation" in the plots).
The spatiotemporal spectral transfers for these simulations are fairly straightforward to understand. When the forcing frequency is not too high, energy and enstrophy are injected at the forcing frequency, and the dissipation ranges are not much affected by changes in the forcing frequency. This allows energy to be dissipated at frequencies lower or higher than the forcing frequency. When the forcing frequency is high, energy and enstrophy are instead injected at relatively lower frequencies.

In general, spatiotemporal spectral transfers show how energy injection varies in wavenumber and frequency simultaneously. Our simulation has very simple spatiotemporal dynamics by design. However, in more realistic simulations, fluids can display significant heterogeneity in wavenumber-frequency space. Figure 12 in \citet{Arbic14} shows wavenumber-frequency plots for transfers by nonlinear advection, where we can see non-trivial spatiotemporal structure at a range of wavenumber and frequencies. Importantly, such spatiotemporal structure cannot be captured simply in plots of either frequency or wavenumber transfers alone. Non-trivial spatiotemporal structure in wavenumber-frequency plots sometimes correspond to identifiable physical phenomena. For example, Figure 2 of \citet{Arro2024} shows wavenumber-frequency plots for magnetic field, velocity, and density of a plasma in a simulation. The plots can be used to identify the presence of phenomena with known dispersion relations, such as Aflv\'{e}n waves. Figure 2 of \citet{torres22} shows wavenumber-frequency plots of various quantities calculated in an ocean-atmosphere coupled simulation. They show that it is possible to use such plots to distinguish between internal gravity waves, mesoscale frontal motions, and submesoscale frontal motions. Importantly, these cannot be distinguished in plots of just frequency or wavenumber transfers alone, because there are frequencies and wavenumber where internal gravity waves coexist with submesoscale motions. 

\subsection{Evolution of temporal transfers}

\begin{figure}
   \includegraphics[width=\columnwidth]{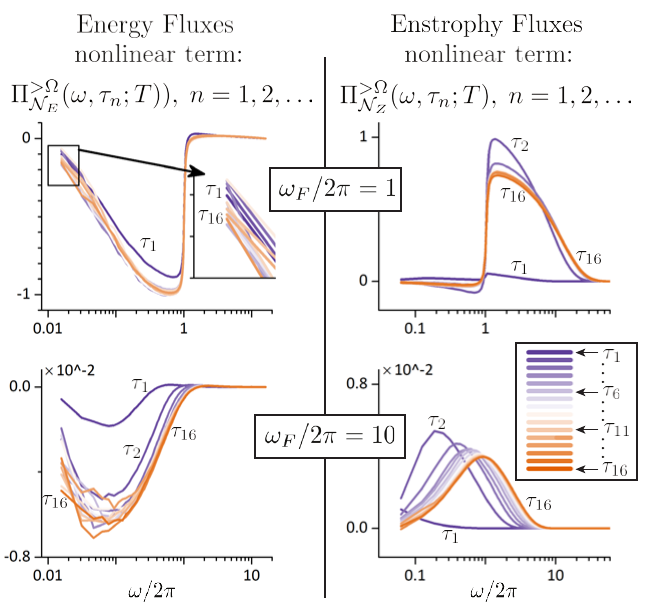}
    \caption{
      Temporal spectral fluxes of the nonlinear advection of energy and enstrophy during spin-up for the $\omega_F/2\pi=1$ and  $\omega_F/2\pi=10$ simulations. The fluxes were calculated for sixteen central times $\tau=\tau_1<\tau_2<\cdots<\tau_{16}$ using contiguous but non-overlapping windows. The window durations were $T=64$ for the energy fluxes and $T=16$ for the enstrophy fluxes.
In both simulations, energy fluxes move to lower frequencies while enstrophy fluxes move to higher frequencies as the system evolves.
}
\label{fig:result_spinup}
\end{figure}

Fig.~\ref{fig:result_spinup} shows how the fluxes due to nonlinear advection of energy and enstrophy evolve as the central time $\tau$ advances during spin-up of the simulations. For each plot, each curve represents nonlinear advection at a distinct central time $\tau$. Purple plots are earlier central times and orange plots are the later central times.
For the simulation with $\omega_F/2\pi=1$, energy and enstrophy are injected at the forcing frequency. From there, energy is transferred to lower frequencies while enstrophy is transferred to higher frequencies. As the simulation spins up, the amount of energy transferred to the lowest frequencies increases somewhat (and similarly for enstrophy at the highest frequencies).
For the simulation with $\omega_F/2\pi=10$, the injection frequency range for both
energy and enstrophy initially develops over a range of frequencies well below the forcing frequency. As the simulation evolves, the injection frequency range
shifts to increasingly higher frequencies.
The dissipation ranges of energy and enstrophy also evolve over time, with the energy dissipation range moving to lower frequencies and the enstrophy dissipation range moving to higher frequencies.  
This evolution of the dissipation ranges corresponds to the development and widening of the spatial cascades of energy and enstrophy.
\subsection{Effects of varying the window size and sampling rate}\label{subsec:results_window}
\begin{figure*}
  \includegraphics[width=0.67\textwidth]{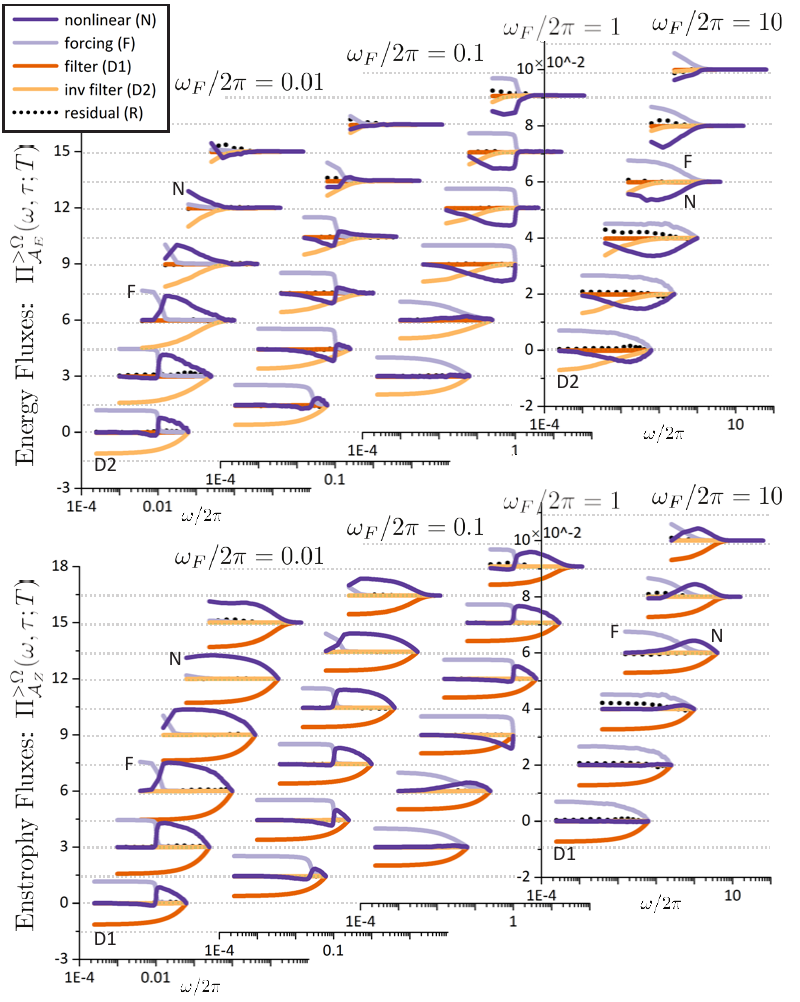}
  \caption{
    The effect of window size ($T$) and sampling rate on temporal spectral fluxes of energy and enstrophy.
    Each column corresponds to fluxes calculated from the same simulation with the same forcing frequency. All fluxes were calculated by sampling 512 snapshots in time from the same simulation. For each column, the bottom set of fluxes samples the ensemble of simulations the least frequently while the top set of fluxes samples the ensemble the most. Just as with other discrete Fourier transforms, the range of frequencies is set by the effect of the window size and the sampling rate. For each vertical axis, the lower-leftmost set of fluxes is plotted correctly centered at zero, while all other fluxes are shifted vertically for ease of comparing fluxes. The rightward tilt occurs because the difference in sampling between plots causes a translation in the output frequency scales of the STFT. The STFT of the simulation output with the lowest sampling rate returns fluxes on the lowest range of frequencies, while the STFT of the simulation output with the highest sampling rate returns fluxes on the highest range of frequencies.}
\label{fig:result_windowingsampling}
\end{figure*}

In practice, the lowest and/or highest frequencies relevant to the dynamics of a fluid might not be resolved by a data set, due to insufficient time series duration or sampling rate. To aid in the interpretation of temporal fluxes used as a diagnostic for real data, we present in Fig.~\ref{fig:result_windowingsampling} a systematic study of the effect of temporal resolution on the resulting temporal fluxes for our two-dimensional simulations. While we discuss the effect of $T$ and sampling rate in Sec. \ref{subsec:spatialtemporaltheory} on the frequency ranges of temporal spectral transfers and fluxes, it is not possible to analytically show their effect on the magnitude of temporal spectral transfers at different frequencies. Thus, the effect of these parameters on spectral fluxes, the integrals of the transfers, is opaque.
This investigation reveals that the temporal fluxes typically give varying but consistent results as the temporal resolution varies, but there are specific cases where the diagnostic fails to give meaningful results.

We show the temporal spectral fluxes $\Pi_{\calA}(\bk,\omega,\tau; T)$ of energy and enstrophy for a wide range of window sizes $T$ and a wide range of sampling rates.
The window sizes were selected from $T\in\{2^{12}, 2^{10}, 2^{8}, 2^{6}, 2^{4}, 2^{2}\}$ 
and the sampling rate was selected such that the total number of snapshots was $512$ for each data set.
Each column corresponds to one of the four simulations (indicated by its forcing frequency).
In each column, the curves in the lowest plot are sampled the least frequently while the curves in the highest plot are sampled most frequently. Each $y$-axis shows the scale for the lower-leftmost plot on that axis.  The remaining plots have been shifted vertically so that they do not all overlap with each other.  Each column  ``leans" to the right, because the sampling rate increases from bottom to top while the number of samples remains constant.  To save space, we overlapped the graphs for the four forcing frequencies, with the graphs for the lowest three forcing frequencies (the left three columns) sharing the same vertical scale.
Considering first the energy dissipation
by the inverse wavenumber filter (yellow curves, labelled ``D2"),
we find that the temporal spectral fluxes are reasonably accurate down to the lowest resolved frequencies.  For simulation output that are too infrequently sampled, the temporal spectral fluxes become inaccurate only at the very highest resolved frequencies, due to aliasing of the unresolved higher-frequency modes into lower modes. 
Fluxes due to the dissipation of enstrophy by the wavenumber filter (orange curves, labelled ``D1") are also accurate at low frequencies, but the aliasing effect at higher frequencies is more apparent.  Despite the aliasing, the total transfer (given by the flux at the lowest resolved frequency) of energy and enstrophy remains roughly the same throughout the simulation output, indicating that aliasing effects are localized to the highest resolved frequencies. Energy dissipation by the wavenumber filter and enstrophy dissipation by the inverse wavenumber filter are negligible for all sampling rates and forcing frequencies.
Considering the forcing term (lavender grey curves, labelled ``F") for both energy and enstrophy, we see that the calculated fluxes are correct if and only if the forcing frequency is not too close to the edge of the interval of resolved frequencies.  For example,  when the forcing frequency is far too low to be resolved by the simulation output with the shortest duration, or when the forcing frequency is far too high to be resolved by the sampling rate, the calculated fluxes develop inaccuracies. When the forcing frequency is near the edge of the interval of resolved frequencies, then the flux due to the forcing tends to decrease to $0$ at too-low frequencies, particularly as especially evident in the case for the energy and enstrophy flux when $\omega_F/2\pi=1$ and $T=2^{10}$ (second set of fluxes from the bottom). When the forcing frequency is slightly lower than the smallest resolved frequency, especially evident in the case $\omega_F/2\pi=0.01$ and $T=2^8$ (third from the bottom), one sees the forcing flux becomes nonzero at higher than accurate frequencies. For most of the simulations, energy and enstrophy budgets have near zero residual flux. We can then determine the effect of the window size and sampling rate on the temporal flux due to the nonlinear advection term (dark purple curves, labelled "nonlinear") by combining the effects on the forcing and dissipation terms.  We note that the nonlinear flux term tends to have good agreement between different sampling rates, \emph{except} when the forcing frequency is barely \emph{un}resolved by the sampling rate, as characterized by the energy fluxes when  $\omega_F/2\pi=1$ and $T=2^{10}$ or by the enstrophy fluxes when  $\omega_F/2\pi=1$ and $T=2^{10}$ or $T=2^8$. Those cases give particularly inaccurate temporal spectral fluxes because neither the forcing frequency nor the enstrophy dissipation range is resolved by the sampling rate. 
These problematic cases may be representative of real-world time series (for large systems, like the ocean, where the shortest time scales are not resolved by data).
However, the fluxes are inaccurate only at the highest resolved frequencies, while the fluxes at the remaining resolved frequencies seem to be reliable.
For both energy and enstrophy, we see slightly non-zero residual fluxes across a wide range of frequencies when  $\omega_F/2\pi=10$ for each of the sampling rates. We also see this for low frequencies in the simulations with the highest sampling rate where $\omega_F / 2\pi = 0.01$ and $1$. Deviations in nonlinear advection between ensembles in that column are explained by these slightly non-zero residual fluxes. We expect the residuals to be non-zero for the instances of highest sampling because these highly sampled simulation outputs do not capture lower frequencies. As a result, aliasing in forcing could lead to a noticeable residual, particularly in the cases where $\omega_F / 2\pi = 0.01$ and $1$ and at the highest sampling rate (top of the first and third columns from the left) because the range of frequencies is slightly higher than the forcing frequencies. A similar thing appears to occur in the farthest right columns where $\omega_F / 2\pi = 10$. For the bottom four plots where $\omega_F / 2\pi = 10$, corresponding to the 4 lowest sampling rates, the residual is likely non-zero where the range of frequencies does not include either the frequency of forcing or the effective frequency of forcing observed in Fig.~\ref{fig:result_fourforcings_w} (bottom left corner plot), which is the frequency at which the simulation responded to the forcing. 
\subsection{Effects of varying the detrending method}
\label{subsec:results_detrending}
For most of our simulations we calculate temporal spectral transfers both with and without a temporal detrending operation (specifically, the removal of a linear trend).  
In previous sections, we reported only on results that did not include a detrending operation.
For time series with sufficiently long duration (Fig.~\ref{fig:result_fourforcings_w}, Fig. \ref{fig:result_fourforcings_transfers_w} and Fig. \ref{fig:result_fourforcings_kw}), because the simulations are statistically stationary, the trends are effectively zero and the detrending operation has
on the temporal fluxes is only seen at the lowest two or three resolved frequencies.

\begin{figure}
    \includegraphics[width=\columnwidth]{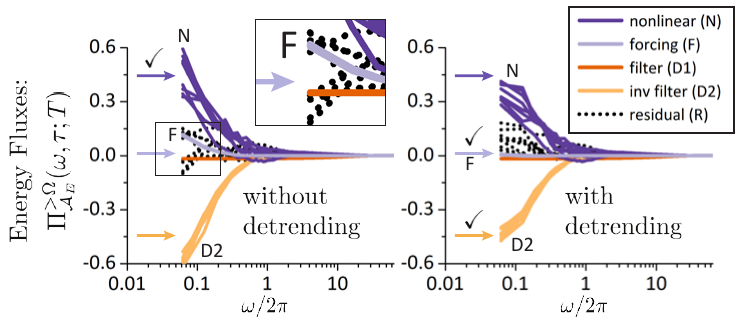}
    \caption{
      Temporal spectral fluxes of energy with and without detrending.
      Each plot shows temporal spectral fluxes for eight independent time series of length $t_{max}=16$ obtained from the lowest frequency ($\omega_F/2\pi=0.01$) simulation. The low frequency forcing effectively provides a trend for these short duration time series. The arrows indicate the ``true" values of the fluxes obtained from a much longer duration time series. Agreements with the ``true'' fluxes, verified numerically,  are indicated by check marks ($\checkmark$).     
 }
\label{fig:result_detrending}
\end{figure}

To study the effect of the detrending operation on the lowest resolved frequencies, we specifically looked at time series that had durations significantly shorter than the forcing period. The relatively low-frequency forcing could be interpreted as the source for a long-term trend.  
In Fig.~\ref{fig:result_detrending}, we analyze  temporal spectral fluxes calculated with or without detrending for eight independent time series of length  $t_{max}=16$ that were forced at the longest forcing period, $2\pi/\omega_F=100$. We retain the same color scheme as previous plots: dark purple curves show nonlinear advection, lavender grey curves show forcing, orange curves show wavenumber filter, yellow curves show inverse wavenumber filter, and black dotted curves show residuals. For comparison, the ``true'' values of the temporal spectral fluxes at the lowest frequencies, determined from a much longer time series, are indicated by arrows.  Detrending brought temporal fluxes for the forcing and inverse wavenumber filter into agreement with the true values for those fluxes.
The flux due to the nonlinear term, however, was not made more accurate by the detrending operation. We note that the residual is significant here only at lower frequencies because short duration datasets cannot capture the low frequency behavior that the full duration dataset contains.

\section{Conclusion}\label{sec:conclusion}

We have shown how temporal and spatiotemporal spectral transfers in fluid turbulence may be defined quite generally in terms of time-frequency analysis methods such as the short-time Fourier transform or the wavelet transform.
Moreover, these methods can be modified to include a fairly general detrending operation in such a way that the temporal and spatiotemporal spectral budgets remain exact even after detrending. 

The interpretation of temporal and spatiotemporal spectral transfers is similar to the interpretation of spatial-spectral transfers, with the main difference being the replacement of time derivative with respect to $t$ by the time derivative with respect to central time $\tau$.
While the spatial spectral transfer $\calA(\bk,t)$ gives the $t$ rate of change of the energy density $\mathcal{E}(\bk,t)$ in wavenumber mode $\bk$, the spatiotemporal spectral transfer $\calA(\bk,\omega,\tau)$ gives the $\tau$ rate of change of energy density $\mathcal{E}(\bk,\omega,\tau)$ in wavenumber-frequency mode $(\bk,\omega)$.
Similarly, the temporal spectral transfer $\calA(\omega,\tau)$ gives the $\tau$ rate of change of the spectral density $\mathcal{E}(\omega,\tau)$ in frequency mode $\omega$.

We derived various theoretical properties of temporal and spatiotemporal spectral transfers, including the effect of a mean flow. We showed that we can decompose the Fourier transform of the nonlinear advection term into asymmetric diagrams, which contain information about the direction of energy transfer, and symmetric diagrams, which fully capture the energetic interaction between any three scales. The diagrams are introduced in this paper to provide a method of tapering the components of the nonlinear advection term. They allow us to develop a nonlinear turbulent triad definition where ringing can be mitigated without changing the useful properties of the triad. Temporal and spatiotemporal transfers of kinetic energy and enstrophy also satisfy detailed conservation laws. We use metrics from turbulence studies of wavenumber triads to quantify the effect of sweeping by a mean flow on the locality of the triad interactions in the frequency domain \cite{zhou93, lesieur1997}. Much of the theory from this section can be used outside of kinetic energy transfers and fluxes, as we show in Appendix \ref{app:triadsofscalars}.
The issue of locality versus nonlocality of triad interactions is particularly important to the interpretation of temporal and spatiotemporal fluxes.  One might be inclined to think of spectral fluxes as arising from a local operator, but in wavenumber, frequency, or wavenumber-frequency space, a flux is almost always a nonlocal function of the spectral quantity of interest.  In three dimensions, this operator is expected to be somewhat narrow in wavenumber, and in two dimensions a bit wider \cite{Kraichnan71}.  However, we have shown how the operator may become significantly wider in frequency due to sweeping by a mean flow.  
Thus, one should keep in mind the relatively greater degree of nonlocality in the frequency domain when diagnosing spectral transfers and fluxes.
Studying the relationship between energy and enstrophy spatiotemporal transfers, one cannot show that energy and enstrophy temporal fluxes generally flow in opposite directions (i.e. the dual cascade) as was shown for spatial fluxes by ~\citet{ kraichnan67}.  Moreover, one can show that the presence of a mean flow 
 and high wavenumber modes can change the locality of triad interactions. Local interactions can be delocalized due to a constant sweeping velocity.
 For these reasons, the existence of a dual cascade in frequency space seems less likely than its existence in wavenumber space.
 However, the temporal and spatiotemporal triad interactions should not be discounted as being fully determined by sweeping, because only the triad interactions involving high wavenumbers are significantly affected by sweeping, and the low-wavenumber, low-frequency modes of the flow are also of dynamical interest.
 To illustrate the use of temporal and spatiotemporal spectral transfers and fluxes,
 we applied the diagnostic to the output of numerical simulations of two-dimensional turbulence.
These simulations show how the frequency affects the transfers of energy and enstrophy in both wavenumber and frequency space.
We also used the simulations to
examine how the diagnostic is affected by insufficient window size, insufficient sampling rate, and the use of detrending, which are important for practical applications.
The main physical result of the numerical investigation is that temporal energy  due to the nonlinear advection term can move to larger or shorter timescales relative to the scale of injection. In other words, it is possible to inject energy at a wide range of frequencies, including frequencies above and below the energy dissipation range.  The lowest forcing frequency case  ($\omega_F/2\pi=0.01$)  is particularly interesting, as it shows energy may be injected at very low frequencies and then transferred to higher dissipation frequencies, indicated by positive fluxes in Fig. \ref{fig:result_fourforcings_w}.  This result may be related to the positive temporal energy flux calculated by ~\citet{Arbic12,Arbic14} for one region of the ocean.  In that one region, the time scale associated with energy injection by baroclinic instability is much longer than the eddy time scale, significantly more so than in the other regions considered. However, there may be other explanations, such as inadequate spatiotemporal resolution of the data, for the positive temporal energy flux observed  in ~\citet{Arbic12, Arbic14}. We have shown here that a positive flux of energy in two dimensions is possible when energy is injected at low frequencies. Enstrophy seems to only transfer to shorter timescales.

We observed that if the forcing frequency is increased beyond some threshold, the energy and enstrophy injection occurs at some lower frequency that remains below the high frequency dissipation range. It is yet unclear what determines this threshold for either the energy or the enstrophy.

Unlike the temporal fluxes, the spatial energy and enstrophy fluxes are minimally affected by changing the forcing frequency, except for a decrease in the magnitude of the flux for high forcing frequency. At the highest forcing frequency, the decrease in flux magnitude is due to the decrease of the total energy and enstrophy injected into the system.

The main practical result of the numerical investigation is that the temporal and spatiotemporal spectral transfers and fluxes for finite time series are quite robust to limitations typical of realistic data. Changes to window size, sampling rate, and detrending method usually have little effect on the temporal spectral transfers or fluxes except at marginal frequencies.  For short duration time series, temporal spectral fluxes tend to be accurate even when the lowest frequencies in the system are not resolved by the data.  When the highest frequencies are not resolved, there is aliasing into the highest resolved frequencies, but the temporal spectral fluxes at the remaining lower frequencies remain accurate.

We found that one must be particularly careful in the special case where the forcing frequency is close to -- either above or below -- the highest resolved frequency. In this case, the temporal spectral fluxes become particularly untrustworthy at nearby frequencies. The worst effect, fluxes having the wrong sign, was seen when the forcing frequency was slightly too high to be resolved and the dissipation range was also not fully resolved. One should be skeptical of the highest resolved frequencies in any case due to the Heisenberg-Gabor uncertainty principle.

Our investigation also showed that when the forcing frequency is too low to be resolved, a linear detrending operation may increase the accuracy of the temporal spectral transfers at low frequencies. This effectively tells us that detrending temporal transfers can be an effective method of dealing with nonstationarity in time series. The accuracy was improved for all of the temporal spectral transfers except for the transfer due to nonlinear advection. The accuracy of the nonlinear transfer was worsened by detrending. For this reason, it would be reasonable to diagnose temporal spectral transfers both with and without detrending. 

We find that our STFT-derived spatiotemporal transfers provide a relatively robust diagnostic for studying spatial and temporal scales in fluid dynamics. While the diagnostic has been used primarily in idealized simulations in the past, we expect this method to be useful for studying more realistic, high-resolution simulations and observational data. In particular, \citet{skitka23} demonstrates the usefulness of spatiotemporal diagrams in precisely tracking the source and destination scales of energy transfer in simulations. Given the broad applicability of spatiotemporal transfers and triads, we expect that the \citet{skitka23}'s use of diagrams can be generalized to studies of cascades outside of kinetic energy, such as the temperature variance studied by \citet{Martin2021}. We discuss such diagrams in Appendix~\ref{app:triadsofscalars}. 

We briefly use the diagnostic to study the effect of sweeping by a constant flow on spatiotemporal transfers. Based on this analysis, we expect the spatiotemporal diagnostic to be useful for numerically probing models of isotropic random sweeping, discussed by \citet{Tennekes75, wilczek2012, chen89}. However, models of sweeping by constant flow will be even more useful for theoretical and numerical research that probes sweeping on kinetic energy transfers and transfers of other tracer budgets. This is especially important for understanding how to interpret spectra derived from satellite data in oceanographic or solar wind studies. Pending questions about the method aside, this work has demonstrated the potential for broad application of this diagnostic to fluids in a variety of contexts. In particular, we anticipate that as observational data and simulations increase in resolution, there will be increased value in using temporal and spatiotemporal transfers to better understand interactions between different scales. Research focused on studying the effects of submesoscale and mesoscale ocean features (like eddies) can take advantage of the framework we introduce in this paper to better understand their effect on large scale currents. \citet{Martin2021} and \citet{Hochet2020} show that the framework discussed in this paper can be used to study the transfers between different time and spatial scales of temperature variance in the ocean or ocean-atmosphere interface. Similar work can be carried out with higher resolution simulations of the climate to better understand air-sea interactions relevant to the weather or climate.

Magnetohydrodynamics (MHD), the study of the motion of charged fluids such as plasmas, is another field where the effects of turbulence are studied at different scales. Transfers in space have been studied in theory and simple simulations using spectral  transfers \cite{Dar2000, Danilov2001, verma2005}. Timescales are also of interest for astrophysical flows \cite{zhou2004}, but have typically been studied using time correlation methods and Taylor's hypothesis \cite{Matthaeus2010, Servidio2011, Matthaeus2016}. Correlations are often useful for relatively sparse time series data. However, we have demonstrated in this paper that our diagnostic can robustly characterize the spectral behavior of turbulent flows in frequency space even with sparse data. We assert that our diagnostic could be useful for analyzing satellite data of plasma flows, as studied in \citet{Matthaeus2016}. One specific MHD problem that can be addressed with the diagnostic proposed here is ``1/f", or ``flicker" noise \cite{Wang2024}. Such noise manifests in the magnetic field frequency spectra of plasmas emitted from the sun. At large time scales, it is observed that this spectra follows $\omega^{-1}$ scaling. The origin of this scaling is a mystery in plasma physics. One potential way to understand this scaling is to use our frequency transfer diagnostic to understand how magnetic field energy is transferred between scales. One can try to use frequency transfers in data to understand how magnetic field energy is transferred to and from large time scales. One paper that studies the transfer of magnetic field energy to large time scales is \citet{Arro2024}, which uses coarse-graining to derive spatiotemporal fluxes, using the method derived in \cite{sadek2018} for calculating spatial spectral fluxes. They identify magnetic field energy fluxes that move from high to low frequencies. While they refer to these as an inverse cascade, they do not analyze the locality of these transfers of magnetic field energy. Future research could use our spectral diagnostic to gauge scale-locality in frequency and wavenumber to determine whether it is a true ``cascade." Admittedly, this is also possible with the coarse-graining formalism \cite{Aluie_Eyink_09_partI, Aluie_Eyink_09_partII}. However, coarse-graining cannot access individual diagrams accurately; deriving diagrams requires knowledge of the diagnostic in this paper. These diagrams allow one to not only evaluate locality but also to accurately analyze the magnitude of interactions between different scales. This is a useful tool to discover how phenomena extant at very different scales correlate with each other.

We acknowledge that this paper is missing a discussion of particular physical implications of our diagnostic such as universality or scaling in homogeneous isotropic turbulence. To investigate unversality and scaling, we would need a different simulation. Firstly, our simulation is designed to study fluids in an Eulerian frame. \citet{Tennekes75} shows that the best frame to study temporal spectra is the inertial, or Lagrangian frame, where sweeping does not distort the spectra.  Our calculation in Sec.~ \ref{subsec:galileantheory} confirms that in a moving frame, such as the Eulerian frame, spectra derived from our diagnostic would be subject to distortion by a Doppler shift. 

Furthermore, traditional theoretical calculations of frequency scaling suggest that the kinetic energy displays a $\omega^{-2}$ scaling in inviscid simulations where it is assumed that energy transfers between different frequencies are somewhat local. However, the theory in Sec.~\ref{subsec:locality of triads theory} shows that transfers in frequency are quite nonlocal. This implies that $\omega^{-2}$ scaling is likely a weak approximation that works in specific situations but may not be universal. This paper provides the theory of diagrams, which one could use to probe which frequencies exchange energy, no matter how nonlocal the two frequencies. In future work, this could provide a very interesting way to develop a scaling theory that is independent of assumptions of locality. 

Finally, the simulation we use appears to not be able to access a large number of decades of frequency. Fig. \ref{fig:result_fourforcings_transfers_w} shows this clearly. One can estimate the size of the inertial range by looking at the distance (in frequency) between the maximum of the light purple forcing curve and the minimum of the yellow inverse filter curve in the two plots in the left column, which focus on energy transfers. This corresponds to the frequency difference between the forcing input frequency and the dissipation regime of our fluid, where viscous effects dominate. In the top left plot, this difference corresponds to approximately one decade. In the bottom left plot, it is about two decades. Neither of those simulations appears to provide a wide enough frequency to calculate a convincing power law fit. With a simulation that has a larger inertial range in a Lagrangian frame, we could look for a convincing power law fit.

In conclusion, while there are still fundamental questions to answer about the physical implications of our spatiotemporal transfer method, we anticipate that our spatiotemporal transfer method can be used in numerical and observational studies of kinetic energy and other quantities of interest in studies related to oceanography, meteorology, and magnetohydrodynamics.
\section{Acknowledgements}

A.J.M. thanks Robert Deegan, John Boyd, and especially the late Charles Doering for providing feedback on early versions of this research when that material appeared in A.J.M.’s dissertation. A.M. and B.K.A thank Lachlan Astfalck for providing feedback on revised drafts of this manuscript. A.M. and B.K.A thank William Matthaeus and Shane Keating for valuable discussions about plasma physics. A.J.M. and B.K.A. gratefully thank Mike Messina for assistance with cluster computing carried out on the supercomputer provided by the UM Office of Research Cyberinfrastrucuture (ORCI), which was instrumental for running numerical simulations for this paper. A.M., A.J.M., J.S., and B.K.A. acknowledge support from National Science Foundation (NSF) Grant Nos. OCE-0960820, OCE-1351837, OCE-1851164, OCE-2319144, and OCE-2319142; from Office of Naval Research grant N00014-19-1-2712 and N00017-22-1-2576; and from University of Michigan faculty startup funds. B.K.A is grateful for sabbatical support from Australian National University, in particular from Callum Shakespeare, Andy Hogg, and Adele Morrison, during the academic year of 2024-2025. G.R.F. received support from NSF Grant No. OCE-0960826. R.B.S. conducted this work within the France 2030 framework program, the Centre Henri Lebesgue ANR-11-LABX-0020-01.

\section{Author Contributions}
B.K.A., R.B.S., G.R.F., and A.J.M. conceived and designed the project. A.J.M., A.M., and J.S. developed the theory. A.M. prepared the final manuscript with guidance from B.K.A. and A.J.M., based on an earlier manuscript that A.J.M. prepared with guidance from B.K.A., G.R.F., and R.B.S. The numerical experiments were carried out by A.J.M. with a codebase developed by G.R.F. All authors contributed to editing the manuscript.

\appendix
\section{spatiotemporal Phase Evolution}\label{app:phase}

While our focus is on spectral transfers and therefore on taking the real part of the product of Eq.~\ref{eq:tau_derivative_g} and $\widehat{g}^*$,
we provide insight into the imaginary part as well.
Multiplying Eq.~\ref{eq:tau_derivative_g} and taking the imaginary part gives
\begin{equation}
-\thalf i \left( \ghs\partial_\tau\gh - \gh\partial_\tau\ghs \right) + (1-\alpha)\omega|\gh|^2 = \Imag[\widehat{g}^*\widehat{f}].\label{eq:imag_v1}
\end{equation}
If we write $\gh$ in terms of its real amplitude and phase,
\begin{equation}
\gh(\omega,\tau) = g_0(\omega,\tau)e^{i\phi(\omega,\tau)},
\end{equation}
then Eq.~\ref{eq:imag_v1} can be rewritten as
\begin{equation}
\partial_\tau\phi + (1-\alpha)\omega= \Imag[\widehat{g}^*\widehat{f}]/|\gh|^2
\label{eq:temporal_phase}
\end{equation}
when $|\gh|\neq 0$.
Thus, taking the imaginary part tells us about the $\tau$ rate of change of the phase of $\gh$. In the simpler situation where only spatial spectra of $g$ are calculated, the imaginary part tells us about the $t$ rate of change of the phase of $\gh$; this can be shown with a near identical derivation to the one above. 

Eq.~\ref{eq:temporal_phase} takes different forms depending on the choice of $\alpha$, because different basis functions give different meaning to the phase.  When $\alpha=0$ the phase is measured with respect to $t=0$, and when $\alpha=1$ the phase is measured with respect to $t=\tau$.

Consider the case $\alpha=0$, which is the case where the basis functions $e^{- i\omega t}$ do not move with the window.  The evolution of the phase is
\begin{equation}
\partial_\tau\phi = -\omega + \Imag[\widehat{g}^*\widehat{f}]/|\gh|^2.\;\;\;\;\;\;\;\;\;\;\text{(case $\alpha=0$)}
\end{equation}
If the window is very long, then the phase of $\gh$ measured with respect to $t=0$ will change very little as the window advances.  In other words, $\partial_\tau\phi \approx 0$, so
\begin{equation}
0 \approx \partial_\tau\phi = -\omega + \Imag[\widehat{g}^*\widehat{f}]/|\gh|^2.\;\;\;\;\;\;\;\;\;\;\text{(case $\alpha=0$)}
\end{equation}
We get a balance of terms.  The first term is simply $-\omega$, and the remaining terms must approximately add up to $\omega$ so that the total is approximately zero.

In the $\alpha=0$ case described above, the presence of the $\omega$ term may seem somewhat mysterious.
Why should the rate of change of the phase always include an $\omega$ term that must be balanced by the remaining terms?
The situation is clarified by considering the $\alpha=1$ case, where the basis functions $e^{- i\omega(t-\tau)}$ remain centered with the moving window.  
In this case, the evolution of the phase is simply
\begin{equation}
\partial_\tau\phi =  \Imag[\widehat{g}^*\widehat{f}]/|\gh|^2.\;\;\;\;\;\;\;\;\;\;\text{(case $\alpha=1$)}
\end{equation}
If the window is very long, then the phase of $\gh$ measured with respect to $t=\tau$ must change approximately at the rate $\omega$ as the window advances in order to compensate for the moving window.  In other words, $\partial_\tau\phi \approx \omega$, so
\begin{equation}
\omega \approx \partial_\tau\phi = \Imag[\widehat{g}^*\widehat{f}]/|\gh|^2.\;\;\;\;\;\;\;\;\;\;\text{(case $\alpha=1$)}
\end{equation}

By changing the basis functions, we have eliminated the $\omega$ term, instead incorporating it into $\partial_\tau\phi$.
As in the $\alpha=0$ case, the remaining terms still approximately add up to $\omega$.  The choice of $\alpha$ simply changes the meaning of $\partial_\tau\phi$ and helps to clarify the presence of the $\omega$ term.
The remaining terms are invariant concerning the choice of $\alpha$.

To further illustrate the meaning of Eq.~\ref{eq:temporal_phase}, let us consider the Rossby wave example. We can begin by thinking purely about the spatial phase evolution before we think about spatiotemporal phase evolution. The Rossby wave equation is
\begin{equation}
\label{app:rossbywaveeqn}
\partial_t \nabla^2 \psi(\vb{x},t) = -\beta \partial_x \psi(\vb{x},t)
\end{equation}where $\psi(\vb{x},t)$ is the stream function for a two-dimensional incompressible fluid and $\beta$ is the meridional planetary vorticity gradient. Applying a spatial Fourier transform to Eq. \ref{app:rossbywaveeqn} and dividing by $-k$ gives

\begin{equation}
\label{app:rossbyspatialtransform}
k \partial_t \widetilde{\psi}(\vb{k},t) = i \beta \frac{k_x}{k} \widetilde{\psi}(\vb{k},t),
\end{equation}
where $k_x:= \vu{x}\cdot \vb{k}$ is the component of $\vb{k}$ in the x-direction. Eq. \ref{app:rossbyspatialtransform} can be written in the form of Eq. \ref{eq:general_eom} by letting $\vb*{\chi} = \vb{k}$ and
\begin{equation}
\label{app:gforpsi}
g(\vb*{\chi}, t) = k\widetilde{\psi}(\vb{k},t).
\end{equation}
The general solution to Eq. \ref{app:rossbyspatialtransform} is a sum of linear Rossby waves. The Rossby wave solutions take the form
\begin{equation}
\label{app:rossbysoln}
\widetilde{\psi}(\vb{k},t) = \Psi(\vb{k}) e^{-i(\omega_R t + \theta_0)}
\end{equation}
where $\Psi(\vb{k})$ and $\theta_0$ and $\omega_R$ satisfies the Rossby dispersion relation,
\begin{equation}
\omega_R := -\beta \frac{k_x}{k^2}.
\end{equation}
Each Rossby wave has constant amplitude $\Psi(\vb{k})$, so the energy in each mode $\vb{k}$ does not change with time. Accordingly, the spectral budget given by Eq. \ref{eq:spatial_budget} is simply

\begin{equation}
\pdv{t} \frac{1}{2} |g(\vb*{k},t)|^2 = 0
\end{equation}
The corresponding phase evolution would be given by 
\begin{equation}
\pdv{t} \phi(\vb{k},t) = \beta\frac{k_x}{k^2}
\end{equation}
Noting that the phase is $\phi(\vb{k},t) = -\omega_R t -\theta_0$, we see that the phase evolution recovers the Rossby dispersion. Thus, the phase evolution simply gives the contribution to the dispersion relation by each term in the equation of motion. 

Now, if we turn our attention to spatiotemporal phase evolution, we can take Eq. \ref{app:rossbysoln} and apply an STFT. In the limit of an infinitely long time window, the STFT applied to Eq. \ref{app:rossbysoln} gives
\begin{equation}
\label{eqn: stftrossby}
\widehat{\widetilde{\psi}}(\bk,\omega,\tau) = 2\pi\delta(\omega-\omega_R)\Psi(\bk)e^{-i\theta_0}e^{-i\alpha\omega\tau},
\end{equation}
where $\delta(\cdot)$  is the Dirac delta function. With $g(\vb*{\chi},t)$ given by Eq. \ref{app:gforpsi}, this phase is

\begin{equation}
\label{app:phitau}
\phi(\omega,\tau) = -\alpha \omega \tau - \theta_0
\end{equation}
when $\omega = \omega_R$ and is undefined when $\omega \neq \omega_R$. If $\alpha = 0$ then $\partial_{\tau} \phi = 0$, whereas $\alpha = 1$ then $\partial_\tau \phi = \omega$. Substituting Eq. \ref{app:phitau} into Eq. \ref{eqn: stftrossby} gives the Rossby dispersion relation $\omega = \omega_R$ for either choice of $\alpha$. Thus, the contributions to the $\tau$ rate of change of the phase given by the right-hand side of Eq.~\ref{eq:temporal_phase} once again tell us the contributions of the dispersion relation by various terms in $f$. 
\section{Spectral budget for wavelets}\label{app:wavelet}
 Another commonly used bilinear time-frequency representation is the scalogram, which is the modulus squared of the wavelet transform (WT).
The continuous wavelet transform of $f(t)$ is defined by
\begin{equation}
\mbox{WT}_f(\omega,\tau) := \int_{-\infty}^{\infty}{f(t) |\omega|^{1/2}W((t-\tau)\omega) dt},\label{eq:WT}
\end{equation}
where $W(\cdot)$ is the ``mother wavelet," which can be chosen  according to desired properties.  
While we focus mainly on the STFT, we show here that analagous results hold for the wavelet transform. 

We show that the form of the spectral budget (with and without detrending) remains the same if we replace the short time Fourier transform with the wavelet transform.
In this appendix, we redefine the ``hat'' operation to be the wavelet transform:
\begin{equation}
\widehat{f}(\omega,\tau) := WT_f(\omega,\tau),
\end{equation}
where $WT_f$ is defined by Eq.~\ref{eq:WT}.
Applying the WT to Eq.~\ref{eq:general_eom_2} gives
\begin{equation}
  \int_{-\infty}^{\infty}{(\partial_tg(t)) |\omega|^{1/2}W((t-\tau)\omega) dt}
  = \widehat{f}(\omega,\tau).\label{eq:WT2}
\end{equation}
Integrating by parts and
transforming $t$-derivatives into $\tau$-derivatives gives
\begin{equation}
\partial_\tau \widehat{g}(\omega,\tau) = \widehat{f}(\omega,\tau).\label{eq:wt_tau_derivative_g}
\end{equation}
Multiplying Eq.~\ref{eq:wt_tau_derivative_g} by $\widehat{g}^*$ and then taking the real part gives
\begin{equation}
  \thalf\partial_\tau|\gh(\omega,\tau)|^2 = \Real[\widehat{g}^*\widehat{f}_i], 
\end{equation}
which has the same form as the spectral budget Eq.~\ref{eq:temporal_budget_general} based on the STFT with $\alpha=1$.

In practice, time series are typically not detrended before applying a wavelet transform.
However, if we were to apply detrending then we would obtain the same result as for the STFT with $\alpha=1$:
\begin{align}
  &\widehat{\left(\partial_tg\right)_{detrend}}(\omega,\tau) =\widehat{\left[(\partial_\tau + \partial_t)g_{detrend}(t,\tau)\right]}(\omega,\tau)\\
&= \int_{-\infty}^{\infty}dt\;|\omega|^{1/2}W((t-\tau)\omega)(\partial_\tau + \partial_t)g_{detrend}(t,\tau)\\
  &= \partial_\tau \int_{-\infty}^{\infty}dt\;|\omega|^{1/2}W((t-\tau)\omega)g_{detrend}(t,\tau)\\
  &=\partial_\tau \widehat{g_{detrend}}(\omega,\tau).\label{eq:wt_hatofdetrend}
\end{align}
The resulting spectral budget would take the same form as Eq.~\ref{eq:spectral_budget_detrended}.

\section{Wiener-Khinchin Theorem with STFT}\label{app:wkth}
In applied mathematics, the Wiener-Khinchin (or Wiener-Khintchine) theorem states that the Fourier transform of the cross-correlation of two signals is equivalent to their cross-spectrum if the signals are wide-sense stationary \cite{hlawatsch08, Priestleybook, Elipot2009}. In \citet{Chiu1970}, frequency transfers for atmospheric kinetic energy are calculated by applying a Fourier transform to the covariance of the velocity and the time derivative of the velocity. By the Wiener-Khinchin theorem, this quantity should be equivalent to $\partial_t\mathcal{E}$ in a situation where a valid Fourier transform of the velocity exists. However, we can use the STFT to formalize the method of \citet{Chiu1970} such that it can be used in practical applications. We start with the definition of the cross-correlation,
\begin{equation}
\label{eqn: covariance}
(\xi^{*} * \zeta)(\uptau) = \lim_{Y\to\infty} \frac{1}{Y}\int\limits_{0}^{Y} dt \;\; \xi^*(t) \zeta(t+\uptau),
\end{equation}
where we assume some arbitrary time signals $\xi$ and $\zeta$. We use the ``$*$" notation to denote a cross-correlation between two fields. $\uptau$ here is the time lag of a covariance. $Y$ represents the time scale over which we average the product of the time signals. However, this equation is not appropriate in numerical applications because no real time series is infinitely long. To make this theorem applicable to non-stationary data or simulation output, we first need to redefine the cross-correlation: 
\begin{equation}
    \label{eqn: stftcovariance}
    (\xi^{*} * \zeta)[\sigma](\uptau, \tau) = \lim_{Y\to\infty} \frac{1}{Y}\int\limits_{0}^{Y} dt \;\; \xi^*(t, \tau) \zeta(t+\uptau, \tau+\uptau) \sigma(t - \tau)^2.
\end{equation}

Tapering allows us to take the integral from $0$ to $\infty$ even if we do not realistically know what's happening beyond a finite time limit in the data or simulation output. In this integral, we square the taper function because we are tapering two signals. The taper for the $\uptau$-shifted $\zeta$ is not shifted in time by $\uptau$ because we also shift the central time $\tau$ by $\uptau$. The shifts in both $t$ and $\tau$ cancel out, allowing us to write $\sigma^2$. We shift both $t$ and $\tau$ by $\uptau$ so that both $\xi$ and $\zeta$ are the same size, which is a necessity for this version of the Wiener-Khinchin theorem. Both $\xi$ and $\zeta$ should be detrended by a linear function as per the definition of a cross-correlation function. We leave this out of the notation for readability. 

In the standard Wiener-Khinchin theorem proof, we would introduce the inverse Fourier transform of $\xi$ and $\zeta$. Naively, one may expect to apply an inverse STFT here to make a Wiener-Khinchin theorem for numerical applications. However, an inverse STFT would remove the tapering introduced in Eq. \ref{eqn: stftcovariance}. Instead, we suggest the following equation:

\begin{align}
    \label{eqn: invstft}
    \zeta[\sigma](t, \tau; T, \alpha)= \frac{1}{2 \pi }\int\limits_{-\infty}^{\infty} d\omega \; \widehat{\zeta}[\sigma](\omega, \tau; T, \alpha)
    e^{i \omega (t - \alpha \tau)}.
\end{align}

This inverse Fourier transform relates the tapered $\zeta$ in frequency space to the tapered $\zeta$ in time. By maintaining the tapering, we can now analytically evaluate integrals over the entire real domain, which is necessary for calculating a numerically applicable Wiener-Khinchin theorem. 

With Eq. \ref{eqn: stftcovariance} and Eq. \ref{eqn: invstft}, we can now calculate the relationship between tapered transfers and cross-correlations. We define the same $\alpha$ for both fields in the cross-correlation equation. This amounts to decomposing the two fields in the cross-correlation into the same basis. We choose to start from a cross-correlation between time series of the same size by shifting the central time of $\zeta$ to $T+\uptau$.  We can plug in the complex conjugate of Eq.~\ref{eqn: invstft} and the analogous inverse STFT of $\zeta$ into Eq.~\ref{eqn: covariance} and use the integral definition of the Dirac delta function to derive an STFT version of the Wiener-Khinchin theorem:

\begin{widetext}
\begin{align}
\label{eqn: wktheorem}
(\xi^{*} * \zeta)(\uptau, \tau; T, \alpha) &= \lim_{Y\to\infty} \frac{1}{Y} \int\limits_{0}^{Y} dt \;\; \xi^*[\sigma^{i}](t, \tau; T, \alpha) \;\zeta[\sigma^{i}](t+\uptau, \tau+\uptau; T, \alpha)\\
&= \lim_{Y\to\infty} \frac{1}{Y} \int\limits_{0}^{Y} dt \int\limits_{-\infty}^{\infty} d\omega\int\limits_{-\infty}^{\infty} d\omega_2 \; \frac{\widehat{\xi^*}[\sigma](\omega,\tau; T, \alpha)}{2 \pi} \; \frac{\widehat{\zeta}[\sigma](\omega_2,\tau+\uptau; T, \alpha)}{2 \pi} e^{i\omega (t-\alpha\tau)} e^{-i\omega_2 (t-\alpha\tau)} e^{-i\omega_2 \uptau(1-\alpha)}\\
&=\int\limits_{-\infty}^{\infty} d\omega\int\limits_{-\infty}^{\infty} d\omega_2 \Bigg[\lim_{Y\to\infty} \frac{1}{Y}\int\limits_{0}^{Y} dt \; e^{i (\omega-\omega_2) t}\Bigg]  \frac{\widehat{\xi^*}[\sigma](\omega,\tau; T, \alpha)}{2 \pi} \; \frac{\widehat{\zeta}[\sigma](\omega_2,\tau+\uptau; T, \alpha)}{2 \pi} e^{-i\alpha \tau(\omega-\omega_2)}e^{-i\omega_2 \uptau(1-\alpha)}\\
&= \int\limits_{-\infty}^{\infty} d\omega\int\limits_{-\infty}^{\infty} d\omega_2 \; 2\pi \delta(\omega-\omega_2) \frac{\widehat{\xi^*}[\sigma](\omega,\tau; T, \alpha)}{2 \pi} \; \frac{\widehat{\zeta}[\sigma](\omega_2,\tau+\uptau; T, \alpha)}{2 \pi} e^{-i\alpha \tau(\omega-\omega_2)}e^{-i\omega \uptau(1-\alpha)}\\
&= 
\Bigg\{ 
    \begin{array}{ll}
        \:\frac{1}{2 \pi} \int\limits_{-\infty}^{\infty} d\omega \; \widehat{\xi^*}[\sigma](\omega,\tau; T, 0)\; \widehat{\zeta}[\sigma](\omega,\tau+\uptau; T, 0)\;e^{-i\omega \uptau}, & \text{if } \alpha = 0\\ 
        \frac{1}{2 \pi}\int\limits_{-\infty}^{\infty} d\omega \; \widehat{\xi^*}[\sigma](\omega,\tau; T, 1)\; \widehat{\zeta}[\sigma](\omega,\tau+\uptau; T, 1)\;, & \text{if } \alpha = 1
    \end{array}
%
\end{align}
\end{widetext}

For clarity, we represent the potential results of the above calculation in a piecewise manner. If $\alpha = 0$, the cross-correlation resembles an inverse Fourier transform, akin to the standard Wiener-Khinchin theorem. This means that defining the STFT in terms of complex exponential functions centered at $t = 0$ equates the cross-correlation of $\zeta$ and $\xi$ to the inverse Fourier transform of the cross-spectrum of $\zeta$ and $\xi$, where $\uptau$ is the conjugate time variable to $\omega$.

If one chooses to set $\alpha = 1$, the cross-correlation as defined above is equivalent to the integral of the cross-spectrum of $\zeta$ and $\xi$ with respect to $\omega$. This resembles Plancherel's theorem. This result is not surprising. By setting $\alpha = 1$, the STFT's of $\zeta$ and $\xi$ are defined in terms of the basis of complex
exponential functions centered at each of their respective central times. This negates the time lag $\uptau$ between $\zeta$ and $\xi$. These two cases are indistinguishable if $\uptau = 0$. 

We can demonstrate that this version of Wiener-Khinchin theorem is necessary to use Chiu's auto-correlation method of deriving kinetic energy budgets on data or simulation output. They focus on the case of $\uptau = 0$. Even so, without the adjustments due to the STFT, the budgets calculated by \citet{Chiu1970} suffer from the same problems as other spectral budgets calculated in the frequency domain without adjusting for convergence at $t=-\infty$ and $t=\infty$ and non-stationarity of time series. We can calculate the zero-lag ($\uptau=0$) cross-correlation of $g(t,\tau; T,\alpha)$ and $\partial_t g(t,\tau; T,\alpha)$:
\begin{widetext}
\begin{align}
     \bigg(g^{*}(t,\tau; T,\alpha) * \partial_t g \bigg)(\uptau = 0) &= \lim_{Y\to\infty} \int\limits_{0}^{Y} dt \; g^{*}(t,\tau; T,\alpha) \pdv{t} g(t,\tau; T,\alpha)\\
    &= \frac{1}{2\pi} \int\limits_{\omega=-\infty}^{\omega=\infty} d\omega \; \widehat{g}^{*}(\omega,\tau; T,\alpha) \bigg[\pdv{\tau} + i\omega(1-\alpha)\bigg] \widehat{g}(\omega,\tau; T,\alpha) .
\end{align}
\end{widetext}

If we set $g = \vb{u}$ and take the real part of the above equation, we can show the derivation of the $\tau$-derivative of kinetic energy:
\\
\\
\\
\begin{widetext}
\begin{align}
 \Real\bigg[\bigg( \vb{u}^{*}&(t,\tau; T,\alpha) * \pdv{\vb{u}}{t}\bigg)(\uptau = 0)\bigg]=
  \frac{1}{2\pi} \int\limits_{\omega=-\infty}^{\omega = \infty} d\omega \; \pdv{\tau} \frac{1}{2} |\widehat{\vb{u}}(\omega, \tau; T, \alpha)|^2= \frac{1}{2\pi} \int\limits_{\omega=-\infty}^{\omega = \infty} d\omega \; \pdv{\tau} \mathcal{E}
\end{align}
\end{widetext}
One can repeat this above calculation for every other term in kinetic energy dynamics to calculate the full kinetic energy budget using this modified Wiener-Khinchin theorem. 

Additionally, the above calculation gives us insight into how our power spectra, calculated from STFT's, relates to the cross-correlation. Even with tapering, one can relate cross-spectra to cross-correlations. In cases where one is interested in understanding the effects of a lagged correlation, we show here that one needs to use the STFT from Eqn. \ref{eqn: wktheorem} with $\alpha=0$.

\section{Detrending}\label{app:detrend}

To perform a linear least squares fit we minimize the integral of the squared residuals,
\begin{equation}
  S(\tau) \defeq \int_{\tau-T/2}^{\tau+T/2}dt\left|f(t)-\sum_n c_n(\tau)\varphi_n(t-\tau)\right|^2.
\end{equation}
Changing the integration variable to $t^\prime\defeq t-\tau$ gives
\begin{equation}
  S(\tau) = \int_{-T/2}^{T/2}dt^\prime\left|f(t^\prime+\tau)-\sum_n c_n(\tau)\varphi_n(t^\prime)\right|^2.
\end{equation}
To find $\{c_n\}$ that minimize $S$ we solve $\partial_{c_n}S=0$.  The result is
\begin{equation}
\braket{ f(t^\prime+\tau)} {\varphi_n^*(t^\prime)} = \sum_m c_m(\tau)\braket {\varphi_n^*(\tp)}{\varphi_m(\tp)}_{t^\prime}. \label{eq:c_indirect}
\end{equation}
We define the matrix
\begin{equation}
\label{eqn:Mdef}
M_{nm} \defeq \braket {\varphi_n^*(\tp)}{\varphi_m(\tp)}_{t^\prime},
\end{equation}
and note that it is independent of both $t$ and $\tau$ and only depends on the form of the basis functions.
Eq.~\ref{eq:c_indirect} can then be solved for $c_n(\tau)$ using the matrix inverse of $M$, giving the result in the main text, Eq.~\ref{eq:c_n}.
\begin{widetext}
Using Eq.~\ref{eq:trend_explicit} to calculate $(\dot{g})_{trend}$, $\partial_t g_{trend}$ and $\partial_\tau g_{trend}$ we obtain
\begin{equation}
(\dot{g})_{trend}(t,\tau) = \sum_n \varphi_n(t-\tau)\sum_m (M^{-1})_{nm}\braket{ \dot{g}(t^\prime+\tau)} {\varphi_m^*(t^\prime)}_{t^\prime}.
\end{equation}
\begin{equation}
\partial_t g_{trend}(t,\tau) = \sum_n \dot{\varphi}_n(t-\tau)\sum_m (M^{-1})_{nm}\braket{ g(t^\prime+\tau)}{ \varphi_m^*(t^\prime)}_{t^\prime}.
\end{equation}

\begin{align}
  \partial_\tau g_{trend}&(t,\tau) = -\sum_n \dot{\varphi}_n(t-\tau)\sum_m (M^{-1})_{nm}\braket {g(t^\prime+\tau)} {\varphi_m^*(t^\prime)}_{t^\prime}+ \sum_n \varphi_n(t-\tau)\sum_m (M^{-1})_{nm}\braket{\dot{g}(t^\prime+\tau)}{ \varphi_m^*(t^\prime)}_{t^\prime}.
\end{align}
\end{widetext}
Combining these three results gives Eq.~\ref{eq:trendofgdot}.

\section{Generalized Transfers of Tracers}\label{app:triadsofscalars}

In this paper, we primarily study the incompressible Navier-Stokes equations to test the spatiotemporal transfer diagnostic on kinetic energy dynamics. In this section, we examine other examples of equations of motion (i.e. examples of Eq.~\ref{eq:general_eom}). \citet{Martin2021} and \citet{Hochet2020} studied temporal fluxes and transfers of a temperature variance budget, which was derived from the equation for $\partial_t \theta$, where $\theta$ here is the potential temperature of a fluid. We assert that one can study fluxes and transfers of any tracer budget subject to advection by velocity with corresponding dynamics described in a simulation, such as temperature, salt, biomass, etc. Here, we discuss diagrams of tracers that experience nonlinear advection. We can generalize Eq.~\ref{eq:general_eom} by setting $g = \eta$, such that $\eta$ is a tracer quantity that experiences nonlinear advection. Then, the equations of motion can be written as 
\begin{align}
\partial_t \eta(\vb{x},t)& = -(\bu \bcdot\bnabla)\eta + Q_\eta(\vb{x},t),\label{app:advectioneq}\\\;\;\;\;\bnabla\bcdot\bu&=0, \label{app:incompressq}
\end{align} 

All the tapering and detrending operations considered in the main paper can be applied to the above equation. Here, $Q_\eta(\vb{x},t)$ represents all the additional sources or sinks of $\eta$ at a point in space and time. For example, if $\eta$ refers to temperature or a solute concentration, diffusion would represent a sink included in $Q_\eta(\vb{x},t)$. We can convert the above equation of motion to a budget equation by multiplying it with $\eta^*$ and taking its real part:
\begin{align}
\frac{1}{2}\partial_t |\eta(\vb{x},t)|^2 = - \Real\bigg[\eta^*(\bu \bcdot\bnabla)\phi\bigg] + \Real[\eta^* Q_\eta(\vb{x},t)].
\label{eqn:transfersphiQ}
\end{align}
One could refer to this budget as an $``\eta$-variance". $ \Real[\eta^* Q_\eta(\vb{x},t)]$ represents all the non-advective transfers of $\eta$-variance. More explicitly, a positive transfer of $\eta$-variance at wavenumber-frequency mode $(\vb{k}, \omega)$ corresponds to an increase of the variability of $\eta$ at wavenumber $\vb{k}$ and frequency $\omega$. One can also write the flux by nonlinear advection of the tracer flux similar to Eq.~\ref{eq:N}:
\begin{widetext}
\begin{align}
\calN_\eta[\sigma](\bk,\omega,\tau) &:= \Real[\wht{\eta}^*[\sigma](\bk,\omega,\tau)(\wht{\bu\cdot\bnabla\eta})[\sigma](\bk,\omega,\tau;T)]\label{eq:Nphi}\\
& = \sum_{\bp,\bq}\int d\omega_{\bp}\int d\omega_{\bq}\;\mathcal{T}_\eta[\sigma_1, \sigma_2](\bk,\bp,\bq,\omega,\omega_{\bp},\omega_{\bq},\tau;T)\delta_{\bk-\bp-\bq,\bm{0}}\delta(\omega-\omega_{\bp}-\omega_{\bq}),
\end{align}
\end{widetext}
where the diagram $\mathcal{T}_\eta$ is defined as
\begin{widetext}
\begin{align}
\label{app:qtriads}
\mathcal{T}_\eta [\sigma_1,\sigma_2](\bk,\bp,\bq,&\omega,\omega_{\bp},\omega_{\bq},\tau;T)  :=  
\Bigg\{ 
    \begin{array}{cc}
        \:0, \: \: \:  \: \: \: \: \: \: \: \: \:  \: \: \: \: \: \:  \text{       if any of } \omega, \omega_{\vb{p}}, \omega_{\vb{q}}, \vb{k}, \vb{p}, \text{ or } \vb{q} = 0,\vspace{0.2cm}\\ \Imag\bigg[\Big(\widehat{\widetilde{\eta}}^{*}[\sigma_1\sigma_2](\bk,\omega,\tau)\widehat{\widetilde{\eta}}[\sigma_1](\bq,\omega_{\bq},\tau)\Big)\Big(\vb{k}\cdot\wht{\bu}[\sigma_2](\bp,\omega_{\bp},\tau) \Big)\bigg], \: \: & \text{otherwise}.
    \end{array}
\end{align}
\end{widetext}
The tapers defined here follow the same constraints as those in Eq.~\ref{eq:unsym_triad_defined_check}. Unlike with kinetic energy transfer diagrams, these asymmetric tracer transfer diagrams cannot be summed to get a symmetric tracer diagram. Additionally, there is no detailed conservation law for these triads. However, the notion of a unique direction of transfer remains in Eq.~\ref{app:qtriads}: $(\vb{k}, \omega)$ denotes the destination wavenumber-frequency mode; $(\vb{q}, \omega_{\bq})$ is the source. For a $\eta$-variance, a positive value for Eq.~\ref{app:qtriads} would represent a decrease in the magnitude of $\eta$ variability in the $(\vb{q}, \omega_{\bq})$ mode and an increase in the magnitude of $\eta$ variability in the $(\vb{k}, \omega)$ mode. A cascade in these quantities would correspond to variability transferring from an injection scale to either larger or smaller scales in space and/or time. Measures like this would be useful for understanding space and time scales of variability in the dynamics of quantities such as temperature, salinity, or any other tracer of interest related to fluid dynamics. Parseval's theorem can be used to obtain just the temporal or spatial transfers akin to Eq.~\ref{eqn:temporaltransfers}. 
The triads in Eq.~\ref{app:qtriads} and the transfers in $\phi$ all follow the same change in locality caused by sweeping by a constant velocity, namely Eq.~\ref{eq:transformation_rule} for general transfers and Eq.~\ref{eq:triad_galilean} for the triads. Thus, as with kinetic energy transfers, large wavenumber modes and sweeping can delocalize interactions. 

In magnetohydrodynamics (MHD), velocity is not the only vector field responsible for nonlinearly advecting other quantities. The standard equations for magnetohydrodynamics incorporate contributions from magnetic fields advecting magnetic fields, and magnetic fields advecting velocity fields. \citet{Dar2000} and \citet{Dar2001} introduce spatial fluxes and transfers of kinetic \textit{and} magnetic energy in plasma systems. They also introduce symmetric and unsymmetric magnetic energy transfer diagrams. These quantities can be generalized to frequency, as we did for kinetic energy with Eqns. \ref{eq:unsym_triad_defined_check} and \ref{eq:triad_defined_check}. We reserve any analysis of the locality of these tracers for future work, due to the complexity and uniqueness of the equations of magnetohydrodynamics. 

\section{Numerical forcing}\label{app:forcing}

We construct a statistically stationary, statistically isotropic forcing that is narrowband in both wavenumber and frequency. We begin with defining a deterministic forcing amplitude in wavenumber space:
\begin{widetext}
\begin{align}
\label{eqn: wvnbrforcing}
f_0(\bk):=
\Bigg\{ 
    \begin{array}{cc}
        \:\Big[(k_F+\Delta k_F)^2-|\bk|^2\Big]\: \: \Big[|\bk|^2-(k_F-\Delta k_F)^2\Big], & \text{if } | |\bk| - k_F| < \Delta k_F \vspace{0.2cm}\\ 
        0 & \text{otherwise},
    \end{array}
\end{align}
\end{widetext}
where $k_F$ sets the location peak of the forcing amplitude in wavenumber and $\Delta k_F$ sets the range of wavenumber over which the forcing is non-zero.
We choose $k_F =  59$ and $\Delta k_F=4$, which provides roughly one decade each for the development of the energy and enstrophy cascades. 

We then add stochasticity in time and wavenumber using a scheme originally introduced by ~\citet{Lilly69} and slightly modified to be radially symmetric on average in wavenumber by ~\citet{Maltrud91}. This scheme creates a forcing amplitude with a power spectrum centered at $\omega = 0$ that roughly resembles a Lorentzian in frequency space. For each wavevector $\bk$, the forcing amplitude at time-step $n$ is defined in terms of the force at the previous time-step as follows:
\begin{align}
\label{eqn: forcingamplitude}
\widetilde{F}_{0}^\pm(\bk,t_n) = R \widetilde{F}_{0}^\pm(\bk,t_{n-1}) + f_0(\bk)\sqrt{1-R^2}e^{i\phi_n^\pm(\bk)},
\end{align}
where $\{ \phi_n^\pm(\bk) \}_{n=0}^\infty$ is a set of independent, identically distributed random phases on the interval $[0,2\pi)$.  The parameter $R\in[0,1]$ is a dimensionless correlation coefficient that defines an integral time scale of $\tau_{F_0} = 0.5\Delta t (1+R)/(1-R)$,  where $\Delta t$ is length of a single time-step. The case $R=0$ corresponds to white noise forcing, while the case $R=1$ corresponds to constant, nonrandom forcing. The $\pm$ superscript indicates that we are defining two statistically independent forcing amplitudes $\widetilde{F}_{0}^+$ and $\widetilde{F}_{0}^-$. 

To add deterministic periodicity to the forcing (as opposed to the potential stochastic periodicity that could arise from Eq.~\ref{eqn: forcingamplitude}), we define the full forcing as

\begin{align}
\widetilde{F}(\bk,t) :=  \widetilde{F}_{0}^+ (\bk,t)e^{+i\omega_Ft}    + \widetilde{F}_{0}^-(\bk,t)  e^{-i\omega_Ft} .
\end{align}

Here, we introduce $\omega_F$ as a forcing frequency by multiplying $\widetilde{F}_{0}^+$ by $e^{+i\omega_Ft}$ and $\widetilde{F}_{0}^-(\bk,t)$ by $e^{-i\omega_Ft}$.
One can show that the integral time scale of $\widetilde{F}$ is the same as for $\widetilde{F}_0^\pm$ (i.e.~$\tau_F = \tau_{F_0}$).
In order to obtain a sharp spectral peak in the power spectrum of the forcing, the integral time scale 
$\tau_F$
must be set sufficiently large relative to the forcing period $2\pi/\omega_F$.  We choose $\tau_F = 5\times 2\pi/\omega_F$, which produces the relatively narrow spectral peaks shown in Fig.~\ref{fig:forcing}.

\section{Time scales for nondimensionalization}\label{app:nondim}

To remove dimensions from the frequency scale, we identify three options. The first option, inspired by modulated turbulence literature ~\citep{Heydt03}, combines the  mean square velocity of the fluid with the forcing wavenumber:

\begin{equation}
t_{nat} = \frac{c}{\langle|\bu|\rangle k_F/2\pi} \implies \frac{\omega_{nat}}{2\pi} =  \frac{\langle|\bu|\rangle k_F/2\pi}{c},\;\;\; c\gtrsim 1 ,\label{eq:tnat}
\end{equation}
where $c$ is some dimensionless constant, $\langle|\bu|\rangle$ is the average Eulerian speed, and $k_F$ is the forcing wavenumber.

We choose a value of $c$  based on the simulation with $\omega_F/2\pi = 10$.  In the  $\omega_F/2\pi = 10$ simulation, energy and enstrophy are  injected not at the forcing frequency but rather within a range of lower frequencies (as can be seen in Fig.~\ref{fig:result_fourforcings_k} and more easily in Fig.~\ref{fig:result_fourforcings_w}), centered at a frequency which we take to be the natural frequency of the fluid for the forcing scale.
Inserting this frequency into Eq.~\ref{eq:tnat} sets the value $c=2.0$. This is close to the value of $c=2.7$ suggested for modulated three-dimensional turbulence~\citep{Heydt03}.

We introduce a second natural frequency in terms of the locations of the energy and the enstrophy dissipation ranges; in all our simulations, energy is dissipated over a wide range of relatively low frequencies while enstrophy is dissipated over a wide range of higher frequencies. We define the dissipation frequency $\omega_{diss} / 2\pi$ as the frequency that best divides the ranges. Because the edges of the dissipation range in our simulation overlap slightly in frequency, we define $\omega_{diss}^* = 0.63$ based on the geometric mean of the observed range of natural frequencies for the lowest forcing frequencies.   

However, both $\omega_{diss}$ and $\omega_{nat}$ require \emph{post priori} calculations. As an alternative, we define a third dimensionless frequency that can be determined \emph{a priori} using the root mean square amplitude of the forcing defined in Appendix \ref{app:forcing}. The root mean square amplitude of the forcing is calculated to be $\langle |\overline{F(\bx, t)|^2}\rangle^{1/2}  = 8.4$ and has units of frequency squared. This gives a time scale due to the amplitude of the externally imposed forcing:
\begin{align}
t_{ext} &= [\langle |\overline{F(\bx, t)|^2} \rangle]^{-1/4} \approx 0.35 \\ \implies \frac{\omega_{ext}}{2\pi} &= \left[\langle |\overline{F(\bx, t)}|^2 \rangle\right]^{1/4} \approx 2.9.\label{eq:t_ext}
\end{align}

To summarize, the dimensionless frequencies are $\omega/\omega_{nat}$, $\omega/\omega_{diss}$, and $\omega/\omega_{ext}$. We report results using dimensional frequencies and refer the reader to Table~\ref{table:nondim} to obtain the corresponding dimensionless values. We include $\omega_{nat}$ and 
$\omega_{diss}$ in Figs.~\ref{fig:result_fourforcings_w} and \ref{fig:result_fourforcings_transfers_w} as a reference for frequency scales.The non-dimensional values of the four forcing frequencies are $\omega_F/\omega_{nat}\in\{0.005,0.05,0.5,25 \}$, $\omega_F/\omega_{diss}\in\{0.016,0.16, 1.6, 79\}$, and  $\omega_F/\omega_{ext}\in\{0.0034,0.034,0.34,3.4 \}$.

\begin{table*}[!ht]
  \begin{center}
  \def~{\hphantom{0}}
  \begin{tabular}{c||ccc||ccc}
  \toprule
      $\omega_F/2\pi~$  
      & ~$\omega_{nat}/2\pi$~  
      & ~$\omega_{diss} / 2\pi$~ 
      & ~$\omega_{ext}/2\pi$~
      &  ~$\omega_F/\omega_{nat}$  
      &~$\omega_F/\omega_{diss}$
      &~$\omega_F/\omega_{ext}$\\
      \hline \hline
       ~0.01 & 2.0  & 0.63   & 2.9 &  ~~0.005 & 0.016 & ~0.0034\\
       ~0.1 & 2.0  & 0.63  &   2.9 &  ~~0.05 & 0.16 & ~0.034\\
      ~1  & 2.0 & 0.63    & 2.9 & ~~0.5 & 1.6 &  ~0.34\\
       10 & 0.4 & 0.126   & 2.9 &  ~~25 & 79 & ~3.4\\
  \end{tabular}
  \caption[Dimensionless forcing frequencies.]{ The four forcing frequencies $\omega_F$, scaled by three different frequency scales. The frequency $\omega_{nat}$ is determined by combining the root mean square velocity with $k_F$, $\omega_{diss}$ is determined by averaging the frequency scales of the energy and enstrophy dissipation, and $\omega_{ext}$ is determined by the amplitude of the externally imposed forcing.}
  \label{table:nondim}
  \end{center}
\end{table*}


\bibliography{master8}

\end{document}